\documentclass[12pt]{article}
\pdfoutput=1
\usepackage[
top = 2.5cm, 
bottom = 2.5cm, 
left = 2.5cm, 
right = 2.5cm]{geometry}

\usepackage{jheppub}
\usepackage[sort&compress]{natbib}
\usepackage[utf8]{inputenc}
\usepackage{ifpdf}
\usepackage{braket}
\usepackage{graphicx}
\usepackage{caption}
\usepackage{subcaption}
\usepackage{slashed}
\usepackage{feynmf}
\usepackage{epstopdf}
\usepackage{amsmath, amssymb,url,bm,textgreek,upgreek}
\usepackage{mathrsfs}
\usepackage[svgnames]{xcolor}
\usepackage{hyperref}
\usepackage{dsfont}
\usepackage{float}

\usepackage{tikz} 
\usepackage{tikz-feynman}
\usetikzlibrary{shapes,arrows,positioning,automata,backgrounds,calc,er,patterns}
\usetikzlibrary{decorations.pathmorphing}
\tikzfeynmanset{compat=1.0.0}

\hypersetup{linkcolor=DarkBlue,citecolor=magenta}

\makeatletter
\def\@fpheader{\relax}
\makeatother

\def\be{\begin{eqnarray}}
\def\ee{\end{eqnarray}}

\def\tr{\operatorname{tr}}

\def\Vol{\operatorname{Vol}}

\def\tr{\operatorname{tr}}

\def\d{{\rm d}}

\title{On Marginal Operators in Boundary Conformal Field Theory}
\author{Christopher P. Herzog$^{a,b}$}
\author{and Itamar Shamir$^{c}$}

\affiliation{
$^a$ Mathematics Department, King's College London, \\
The Strand, London,  WC2R 2LS, UK \\
$^b$ C. N. Yang Institute for Theoretical Physics, Department of Physics and Astronomy,\\
Stony Brook University, Stony Brook, NY 11794, USA \\
$^c$ SISSA and INFN, Via Bonomea 265, 34136, Trieste, Italy \\
Emails: \href{mailto:christopher.herzog@kcl.ac.uk}{christopher.herzog@kcl.ac.uk},
\href{mailto:itamar.shamir@sissa.it}{itamar.shamir@sissa.it}.
}

\abstract{
The presence of a boundary (or defect) in a conformal field theory allows one to generalize the notion of an exactly marginal
deformation.  Without a boundary, one must find an operator of protected scaling dimension $\Delta$ 
equal to the space-time dimension $d$ of the
conformal field theory, while with a boundary, as long as the operator dimension is protected, one can make up for the 
difference $d-\Delta$ by including a factor $z^{\Delta-d}$ in the deformation where $z$ is the distance from the boundary.
This coordinate dependence does not lead to a reduction in the underlying $SO(d,1)$ global conformal symmetry group of the 
boundary conformal field theory.  We show that such terms can arise from boundary flows in 
interacting field theories.
Ultimately, we would like to be able to characterize what types of boundary conformal field theories live on the orbits of such deformations.  As a first step, we consider a free scalar with a conformally invariant mass term $z^{-2} \phi^2$, and a fermion with a similar mass.  We find a connection to double trace deformations in the AdS/CFT literature.

}

\begin{document}
\maketitle
\setcounter{page}{2}

\newpage
\section{Introduction}

There is good reason to consider boundaries and defects in quantum field theory (QFT).  
They underlie much of the major progress in theoretical physics over the last thirty years.  For example, they are intrinsic to the development of D-branes in string theory and to the construction of the AdS/CFT correspondence.  They are important in understanding many-body entanglement and topological insulators.
Our focus here is on boundaries in conformal field theories.  As fixed points of the renormalization group flow, conformal field theories (CFTs) provide important landmarks in the space of QFTs more generally.  They also have useful specific applications, for example in describing second order phase transitions in certain condensed matter systems.  

Exactly marginal deformations are of special importance in the study of CFTs. A CFT with exactly marginal operators is a member of a family of CFTs parametrized by the associated coupling constants. Boundary conformal field theories (BCFTs) can accommodate new types of exactly marginal deformations. In addition to the standard \emph{bulk} deformations, which are integrals over all space of dimension $d$ operators, BCFT can include a deformation by a dimension $d-1$ operator integrated on the boundary. Such boundary deformations can be interpreted as motion in the space of conformal boundary conditions.  

The goal of this paper is to propose that BCFTs can admit a new kind of \emph{bulk} marginal deformation which has no equivalence in theories without a boundary (at least, insofar as we are committed to preserve unbroken symmetries). Consider a conformal theory defined on a $d$-dimensional flat space with a boundary, and let us pick coordinates $x^\mu = (z,x^i)$ such that the boundary is given by the equation $z=0$. (We work in Euclidean space throughout.) The marginal deformation we would like to study is built out of an operator $\mathcal{O}_{\Delta}$,
\be \label{non_marginal_def}
\lambda \int \d^d x \, z^{\Delta-d} \, {\mathcal O}_\Delta(x)  \ ,
\ee
where $\lambda$ is a dimensionless parameter. The symmetries of a BCFT are given by the bulk conformal transformations which leave the boundary $z=0$ unchanged. These symmetries are isomorphic to the conformal symmetries of the boundary, namely $SO(d,1)$. 
We emphasise that the deformation \eqref{non_marginal_def} is consistent with all the symmetries of a BCFT. 
Further considerations put constraints on $\Delta$.
In order that the perturbation not diverge as we move away from the boundary, we take the operator to be relevant or marginal, $\Delta \leq d$.%
\footnote{Another possible constraint comes from promoting $\lambda z^{\Delta -d}$ to the expectation value of some other operator $Y$.  In order for $Y$ to be above the unitarity bound, one should take $\Delta \leq \frac{d+2}{2}$ (or $\Delta = d$ when $Y$ is the identity). }  
(When $\Delta=d$, \eqref{non_marginal_def} is identical with the usual notion of marginal deformations.)

While the deformation \eqref{non_marginal_def} preserves all the symmetries of a BCFT (at least classically), it does so in an unusual way. In the standard paradigm of boundary field theories, local Poincar\'e symmetry is preserved in the bulk. 
Said another way, the stress-energy tensor $T^{\mu\nu}$ is fully conserved in the bulk, the violation being localized on the boundary, i.e.\ $\partial_{\mu} T^{\mu\nu} = \delta(z)  n^\nu D$, where $n^\mu$ is a normal to the boundary and $D$ is a local operator on the boundary, called the displacement operator. This state of affairs is natural if one wants to couple the theory to a background metric in a diff invariant way, as we often do. In contrast, \eqref{non_marginal_def} leads to a violation of the conservation of the stress-energy tensor in the \emph{bulk}. Consequently, a naive coupling to a metric will be inconsistent with diff invariance. 
To bypass this difficulty one can introduce a conformal compensator $Y$ and replace in \eqref{non_marginal_def} the factor $z^{\Delta-d}$ by $Y^{d-\Delta}$. The resulting effective action $W[g_{\mu\nu},Y]$ is diff$\times$Weyl invariant.
While conservation and tracelessness of the stress tensor $T^{\mu\nu}$ are lost, a dilatation current 
$x_\mu T^{\mu\nu}$ is conserved for a profile $Y \sim 1/z$.

One may prefer the point of view that such position dependent couplings, which explicitly violate the stress-energy tensor conservation in the bulk, should be ruled out. 
It is natural to wonder then if such violation can emerge dynamically, as a spontaneous rather than explicit breaking.
We provide a brief classical argument, starting with a scalar theory for a field $Y(x)$ with a local interaction $Y^n$.  
A relevant boundary deformation can drive the system to a phase where the classical solution
for $Y$ has a profile that scales with a power of $z$.
Moreover, fluctuations about this classical background will experience position dependent interactions through their coupling to the background value of $Y$. 
In the context of $Y^4$ field theory, this generation is sometimes referred to as boundary ordering or the extraordinary transition \cite{Lubensky:1975zz}.%
\footnote{%
Translations in the normal direction 
are already broken explicitly due to the presence of the boundary. Nevertheless, the current is conserved locally in the bulk and can be thought of as an approximate symmetry of the UV, that is when we probe distances which are much shorter than the distance to the boundary. In this sense, the space dependent vacuum configuration of $Y$ constitutes a spontaneous breaking of this approximate symmetry. }

Just as in conventional CFTs, a marginal deformation typically develops a beta function due to quantum corrections, which leads to a breakdown of conformal symmetry. In other words, a marginal deformation is seldom exactly marginal. The same is expected to be true here. 
In special cases, such as when the dimension is protected by a symmetry, such deformations can give rise to exactly marginal parameters. In this paper we focus on two examples which are guaranteed to preserve conformal symmetry, namely free theories. Specifically, we consider a scalar in arbitrary dimensions with the Lagrangian
\begin{align} \label{scalar_conf_mass}
\mathscr{L}_\phi =  \frac{1}{2} (\partial \phi)^2  + \frac{1}{2} \frac{ \left( \mu^2 - \frac{1}{4} \right)}{z^2}\phi^2 \ .
\end{align}
For not so mysterious reasons we shall call this deformation a \emph{conformal mass}. We will see later why $\mu^2- \frac{1}{4}$ is a useful parameterization of the marginal coupling. In the same way, we can consider a fermion Lagrangian given by
\begin{align} \label{}
\mathscr{L}_\psi = \chi^\dagger \slashed{\partial} \psi - \frac{\mu}{z} \,\chi^\dagger \psi \ .
\end{align}
We consider also in even dimensions the mass term $i \chi^\dagger \gamma \psi$ where $\gamma$ is the chirality matrix.\footnote{%
In a Euclidean setting in 4d, we can take the spinor $\psi$ and conjugate spinor $\chi^\dagger$ to be unrelated.  The relative signs, and factor of $i$ in the case of a $i \chi^\dagger \gamma \psi$ type coupling, are chosen to get a $p^2 + m^2$ type dispersion relation.
}

The deformations we consider have an intriguing affinity to physics in AdS \cite{Carmi:2018qzm,Buhl-Mortensen:2016jqo}. In the Poincar\'e patch, the metric of AdS takes the form $g_{\mu\nu} = \frac{L^2}{z^2}\, \delta_{\mu\nu}$ where $L$ is the AdS radius, and is related to flat space by a Weyl transformation. Performing this Weyl transformation on the Lagrangian in \eqref{scalar_conf_mass}, the factor of $z^{-2}$ in the conformal mass is precisely offset leading to
\begin{align} \label{}
\frac{1}{2}g^{\mu\nu} (\partial_\mu \phi) (\partial_\nu \phi) + \frac{1}{2L^2} \left(\mu^2 - \frac{1}{4} - \frac{d(d-2)}{4} \right) \phi^2 \ .
\end{align}
In other words the conformal mass of flat space is transformed to a standard massive theory in AdS. Unlike \eqref{scalar_conf_mass} the theory on AdS is not conformal, but since the scaling factor $z^{-2}$ no longer appears in the mass term, the symmetries include the full group of isometries of (Euclidean) AdS, namely $SO(d,1)$. These symmetries are the same as of BCFT, except that they are realized in a different way. We will see below that many of the familiar features of physics in AdS, and indeed some aspects of holography, will have parallels in the flat space system describing a particle with conformal mass.

We begin an exploration of the orbit of field theories parametrized by $\lambda$ by computing an effective action $W(\lambda) = - \log(Z(\lambda))$ where $Z(\lambda)$ is the partition function.  
The addition of the conformal compensator $Y$ allows us to generalize the setting of this marginal deformation.  We expect the effective
action  $W[g_{\mu\nu}, Y]$ to be invariant under Weyl transformation up to anomalies.  Instead of considering the flat half-space ${\mathbb R}^+ \times {\mathbb R}^{d-1}$ where this partition function has an IR divergence associated with the infinite volume, we could equally well consider a different conformally flat geometry.  We discussed already the Poincar\'e patch of AdS.  Two interesting examples with finite volume are a hemisphere $HS^d$ (with radius $L$) and a ball (with radius $r<L$) in flat space.

If we start with a profile $Y \sim z^{-1}$ in flat space, on the hemisphere the only dimensionful parameter is the radius $L$, which is introduced through the Weyl transformation.  Despite the existence of a single scale $L$, we find quite generally that the partition function $W(\lambda)$ has a logarithmic singularity $\log(L / \epsilon)$ where $\epsilon$ is a short distance regulator.  
This log divergence is very similar in its behavior to the $a$-type trace anomaly in the stress tensor.  Recall that the $a$-type anomaly is proportional to the Euler density, $T^\mu_{\; \mu} = a {\mathcal E}$ where $a$ is the $a$-anomaly (in an unconventional normalization) and ${\mathcal E}$ is the Euler density, which requires an even dimensional manifold for its definition.  Indeed, when the hemisphere is odd dimensional, its boundary is an even dimensional sphere, and the log divergence of our $W(\lambda)$ is associated to an $\epsilon$ cut-off near the boundary.  In contrast, for the even dimensional hemisphere, in the free scalar and fermion cases where we can calculate $W(\lambda)$ explicitly, 
$\epsilon$ instead comes from a short distance cut-off associated with bringing two operators together, and the logarithm has a bulk origin.

One can think about the log divergence as an anomaly in the scaling symmetry, which is preserved, even though the stress tensor classically is not traceless except in the limiting case $\Delta = d$.  
The usual lore asserts that $a$-type anomalies are independent of marginal couplings because of Wess-Zumino consistency \cite{Osborn:1991gm}.  In contrast, we find explicit dependence on $\lambda$.  The difference comes from the presence of the boundary as we will explain in greater detail in \cite{toappear}.

These Weyl transformations connect computations of the effective action 
$W$ performed in a variety of contexts.  The computations of $W$ in hyperbolic space with a spherical boundary \cite{Gubser:2002zh,Gubser:2002vv,Allais:2010qq} (in the context of double trace deformations) are equivalent to the computations of $W$ on a hemisphere that we discuss here.  
The 
 $HS^4$ partition function for a free scalar ($\mu^2 = 1/4$) was discussed as an example in \cite{Gaiotto:2014gha} where the author proposes more generally a constant term $F_\partial$ in the $HS^4$ partition function as a measure of irreversibility of renormalization group flow for boundary CFTs. 
In the double trace context, ref.\ \cite{Gubser:2002zh,Gubser:2002vv} already pointed out certain monotonicity properties of $W$ under RG flow which were connected to the change in the $a$-anomaly coefficient of a dual even dimensional CFT living on the boundary.  Still in the context of double trace deformations, 
later authors (see e.g.\ \cite{Aros:2011iz}) extended these remarks to the constant $F$ term in the $S^3$ partition function for a CFT living on a three dimensional boundary.  
A Weyl transformation thus relates monotonicity of $F_{\partial}$ with monotonicity of 
$F$ for a class of conformal field theories with holographic duals.\footnote{%
See refs.\ \cite{Rodriguez-Gomez:2017aca,Rodriguez-Gomez:2017kxf} for more recent partition function calculations of free scalars and fermions on these types of spaces.
}

The contents of the paper are as follows.  
In section \ref{sec:symmetries}, we discuss the symmetries that are preserved by our BCFT in the presence of the marginal deformation (\ref{non_marginal_def}).  
In section \ref{sec:emergence}, we show how a conformal mass for the scalar can arise naturally in an interacting scalar field theory with a $Y^{n}$ type coupling and a boundary.
In section \ref{sec:derivative}, we use conformal perturbation theory to determine the response of the partition function to the deformation (\ref{non_marginal_def}) on a ball and hemisphere.  
We find quite generally that in odd dimensions, there is a log divergence in these expressions, corresponding to the fact that the boundary is even dimensional and can support a scale anomaly.  Then in sections \ref{sec:scalar} and \ref{sec:fermion} we consider free scalars and fermions in the presence of a conformal mass.  We will compute the one-point functions $\langle \phi^2 \rangle$ and $\langle \bar \psi \psi \rangle$ and also the hemisphere partition functions.  The hemisphere partition function we compute by two independent means, integrating the coefficient of the one-point function and also diagonalizing the equations of motion on the hemisphere and summing the eigenvalues.
There is a log divergence in $W$ in every dimension which in even dimension comes from a log divergence in the coefficient of the one-point function.
Appendix \ref{app:maps} contains some details of the conformal maps that relate flat space, the hemisphere, and the ball.
Appendix \ref{app:tricks} computes an integral necessary for extracting the log divergence from an integral over the one-point function.

\subsection{Comment about Interactions}

The theories we focus on are free, and the reader may wonder whether this type of deformation remains marginal in the presence of interactions.  
There are in fact several instances in the literature where this type of perturbation to a boundary conformal field theory appears in interacting theories.
In an AdS/CFT context, where the field theory dual is strongly interacting, one finds
 a large collection of so-called Janus solutions in AdS/CFT where a codimension one defect conformal field theory is dual to a higher dimensional gravity system.   
 In these kinds of set-ups, because of supersymmetry there are typically operators with protected dimension.  
 Moreover, one finds that the gravity solution naturally encodes sources for these operators of precisely the form $g z^{\Delta-d}$
 that preserves the residual boundary CFT $SO(d,1)$ symmetry group.  For example, 
ref.\ \cite{Bobev:2013yra} describes a set of 3d boundary conformal field theories (associated with the IR fixed point of maximally supersymmetric 3d Yang-Mills theory) with gravity duals and with scalar operators of protected dimension one and two.  The gravity solutions encode sources for these operators that scale as $z^{-2}$ and $z^{-1}$ respectively, where $z$ is the distance from the defect.  See also ref.\ \cite{Gutperle:2012hy} for an earlier bottom-up model of the same effect.

A second intriguing conventional AdS/CFT example is 
in the work \cite{Horowitz:2014gva}.  The authors consider a strongly coupled field theory with a gravity dual and a conserved current $J^\mu$.  The current couples to an external vector potential with expectation value $A_\nu = \frac{\mu \delta_{\nu 0}}{r}$ where 
 $\frac{\mu}{r}$ can be interpreted as a spatially varying chemical potential.  Such an example is a defect theory analog of the boundary perturbation we have been discussing so far.  The gravity solutions dual to this field theory construction involve interesting hovering black holes in $AdS_{d+1}$.

A third example of a marginal deformation (without $z$ dependence) involves the photon kinetic term in mixed dimensional quantum electrodynamics (QED) \cite{Herzog:2017xha}.\footnote{%
Mixed QED has a large although perhaps little known footprint in the literature.  See refs.\ 
\cite{Gorbar:2001qt, Reystalk, Kaplan:2009kr, Hsiao:2017lch, Hsiao:2018fsc,  Teber:2018jdh, DiPietro:2019hqe}
for several examples where these theories appear.
}
In these theories,  the 4d photon couples to charged 3d matter on the boundary.
The photon kinetic term is an operator with protected dimension $\Delta=4$ and leads to a family of related boundary CFTs,
with a tuneable complexified gauge coupling.
There exist ${\mathcal N}=1$ and ${\mathcal N}=2$ supersymmetric variants of this mixed dimensional QED as well
where the photon kinetic term is generalized to include kinetic contributions from the superpartners 
\cite{Herzog:2018lqz}.

\section{Symmetries and Curved Space}
\label{sec:symmetries}

The ${\mathbb R}^+ \times {\mathbb R}^{d-1}$ in which our boundary CFT lives is related by a Weyl rescaling of the metric to a number of other interesting spaces.  If we can compute correlation functions of primary operators in one of these spaces, then the correlation functions in the other spaces are related trivially via Weyl rescaling.  
Three interesting examples are a spherical boundary in flat space, a hemisphere $HS^d$, and hyperbolic space $H^d$.   
Most straightforwardly, the upper hyperbolic plane can be written with line element
\be
\d s^2 = \frac{L}{z^2} \left( \d z^2 + \delta_{ij} \d x^i \d x^j \right)\ , 
\ee
with $z>0$, 
and is related to flat space via Weyl factor $\Omega(x) = \frac{L}{z}$.  The parameter $L$ is the radius of curvature of $H^d$.  The symmetry group $SO(d,1)$ manifests itself as isometries of the $H^d$ space.  An especially nice feature of the transformation to hyperbolic space is that the explicit $z$ dependence disappears from the perturbation to the action,
\be
\label{AdSdef}
\delta S \to \frac{\lambda}{L^{d-\Delta}} \int \frac{\d^d x}{z^d} {\mathcal O}_\Delta(x) \ .
\ee
(The measure factor $\frac{\d^d x}{z^d}$ is the volume element on $H^d$.)
A number of interesting connections with the AdS/CFT literature emerge from this Weyl rescaling.

We review the details of the maps to the spherical boundary and the hemisphere in Appendix \ref{app:maps}, but the results are easy to state.  The equivalent perturbation for the spherical boundary with radius $L$ is 
\be
\label{balldef}
\delta S \to \lambda \int \d^d x \, \frac{{\mathcal O}_\Delta(x)}
{
\left(
\frac{|r^2 - L^2|}{2 L} 
\right)^{d-\Delta}
} \ ,
\ee
where we could integrate over either $r>L$ or $r<L$ depending on where we want to define the boundary CFT.
For the hemisphere with polar angle $\theta$ and radius $L$ we obtain instead
\be
\label{HSdef}
\delta S \to \lambda \int_{HS^d} \ L^d \d V(S^d)  \frac{{\mathcal O}_\Delta(x) }{(L \cos \theta)^{d-\Delta}} \ ,
\ee
where $\d V(S^d)$ is the volume form on the $d$ dimensional unit sphere and we integrate over the northern hemisphere $0 < \theta < \frac{\pi}{2}$.

We would like to discuss how the $SO(d,1)$ symmetries are realized once the theory is promoted to curved space.
As discussed in the introduction, the space dependence of our marginal deformation is an obstruction to diff invariance. Therefore, promoting the theory to a curved space requires the introduction of conformal compensator $Y$.
Viewing the action as a functional of the external fields $g_{\mu\nu}$ and $Y$, we define the conjugate operators $T^{\mu\nu}$ and ${\mathcal O}$ 
\begin{align} \label{}
T^{\mu\nu} = -\frac{2}{\sqrt g} \frac{\delta S}{\delta g_{\mu\nu}} \ ,
\qquad
\mathcal{O} = -\frac{1}{\sqrt g} \frac{\delta S}{\delta Y} \ ,
\end{align}
where $T^{\mu\nu}$ is the stress tensor.  
(Note ${\mathcal O}$ is related to ${\mathcal O}_\Delta$ of the introduction by a power of $Y$.)
The effective action $W[g_{\mu\nu},Y]$ is invariant under the combined diff and Weyl transformations 
\begin{align} \label{g_Y_var}
\delta g_{\mu\nu} = D_\mu \epsilon_\nu + D_\nu \epsilon_\mu+ 2 \sigma g_{\mu\nu} \ ,
\qquad
\delta Y = \epsilon^\mu \partial_\mu Y - \sigma Y \ . 
\end{align}
This invariance leads to the classical Ward identities
\begin{align} \label{Ward_identity}
D_\mu T^{\mu\nu} = (\partial^\nu Y) \mathcal{O} \ , 
\qquad
T^{\mu}{}_\mu = Y \mathcal{O} \ .
\end{align}

We now restrict to conformally flat geometries, i.e.\ metrics that can be obtained by Weyl rescaling a flat metric, $\delta_{\mu\nu} \to \Omega^2 \delta_{\mu\nu}$. The symmetries correspond to choices of $\epsilon^\mu$ and $\sigma$ in \eqref{g_Y_var} which leave the metric invariant. For conformally flat geometries they are in one to one correspondence with the conformal symmetries of flat space. For each $\epsilon^\mu$ we can construct the current
\be
J^\mu = \epsilon_\nu T^{\mu\nu} \ ,
\ee 
for which it follows from the Ward identities \eqref{Ward_identity} that
\begin{align} \label{}
D_\mu J^\mu = \mathcal{O} \, \delta Y \ .
\end{align}
That is, the current is conserved for choice of $Y$ such that the variation in \eqref{g_Y_var} vanishes. When we consider the full (bulk) conformal group $SO(d+1,1)$ the only possible solution is $Y=0$. However, for theories with a boundary, only conformal Killing vectors $\epsilon^\mu$ for which the normal component vanishes on the boundary are preserved. This leaves the subgroup $SO(d,1)$. In this situation there is a non-trivial solution of $\delta Y=0$.

For convenience, we 
can establish $Y$ by working in the flat half space ${\mathbb R}^+ \times {\mathbb R}^{d-1}$ with coordinates $(z,x^i)$. The boundary leaves the following conformal Killing vectors as symmetries of the action, 
\be
\label{Killingvectors1}
{\rm translations:} \; \; \;  \epsilon^\mu &=& (0, a^i) \ , \; \; \; 
{\rm rotations:}  \; \; \; \epsilon^\mu = (0, \omega^i_{\; j} x^j) \ , \\
{\rm dilations:}  \; \; \; \epsilon^\mu &=& \lambda (z,  x^i)  \ , \\
{\rm special:}  \; \; \; \epsilon^\mu &=& (2 z x^j b_j, 2 x^i x^j b_j - (z^2 + x^j x_j) b^i) \ .
\label{Killingvectors2}
\ee
where $a^i$, $b^i$, ${\omega^i}_j$, and $\lambda$ are constants.  Provided we choose $Y = \frac{1}{z}$, they remain symmetries of the deformed theory as well.  

Having established $Y$ in the flat space case, we can then obtain $Y$ in all the related conformally flat cases by Weyl rescaling.  
From inspection of (\ref{AdSdef}), (\ref{balldef}), and (\ref{HSdef}), we obtain 
\begin{align} \label{}
Y = \begin{cases} 
1 \, , & \text{hyperbolic space}, \\
\frac{1}{|L^2 - r^2|} \, , & \text{spherical boundary}, \\
\frac{1}{\cos \theta} \, , & \text{hemisphere}.
\end{cases}
\end{align}
%
The symmetries follow as well, from Weyl rescaling of the conformal Killing vectors.  In hyperbolic space, the conformal Killing vectors become actual Killing vectors, and the symmetry is realized as the isometry group.

We will see that although scale transformation $x^i \to \lambda x^i$ in flat space and its curved space cousins are classical symmetries of our theories, the partition function in general will have scale dependence.  This scale dependence is a reflection of an anomaly, with many similarities to the standard trace anomaly one finds in typical conformal field theories.

\section{The Emergence of a Conformal Mass}
\label{sec:emergence}

One might want to start with the philosophy that for the UV or microscopic theory, the interactions should not know about the boundary, that initially, there should be no additional position dependent factors of $z$ in the Lagrangian.  
An objection to these position dependent couplings is that they open a Pandora's box.\footnote{%
 We would like to thank O.~Aharony for discussion about this issue.
}  In addition to the marginal operators we discussed in the introduction, there are a plethora of relevant operators as well that we could consider, involving for example powers of $z$ that are larger (less negative) than those required for marginality.  The usual philosophy behind finding a conformal fixed point is that the number of relevant operators is finite, and that tuning them to zero is a feasible process, by adjusting a few external parameters in the system, for example temperature and magnetic field.  With an infinite number of possible relevant operators, finding a critical system becomes much more difficult.

A better philosophy, perhaps, is then that these position dependent couplings should emerge dynamically, from a system where the couplings initially are position independent, and that the rules for such emergence produce a finite number of terms.  
We claim that one such example of emergence is well known in the context of surface critical phenomena, and was studied in detail over thirty years ago.  (See refs.\ \cite{PTandCPreviews,binder1983critical,Diehl:1996kd} for reviews.)  
We generalize their analysis of $\phi^4$ field theory, which is relevant for studying magnetic phase transitions. 
Consistent with the initial absence of position dependent interactions, let us consider the following scalar field theory on a half space:
\be
\label{PhiL}
\mathscr{L} = \frac{1}{2} (\partial Y)^2 +\frac{1}{2}  \lambda^{2n} Y^{2n+2} \ .
\ee
We call the field $Y$ to suggest that it may be a way of giving dynamics to the conformal compensator field $Y$ in the introduction,
at least in the special cases where $n$ leads to a classically marginal operator, e.g.\ $n=1$ in 4d or $n=2$ in 3d.
We keep $\lambda$ very small in order to trust this classical or mean field analysis.
Such a field theory (\ref{PhiL}) admits a classical scaling solution of the form 
\be
\label{Phiprofile}
Y_0 = \frac{1}{\lambda} ( n z)^{-1/n} \ .
\ee

Whether such a solution for $Y$ is actually admissible depends of course on the boundary conditions.  
In the context of this weakly coupled or classical analysis, one can include on the boundary a relevant operator of the form
\be
\mathscr{L}_{\rm bry} = \frac{m}{2}  Y^2  \delta(z)\ .
\ee
A linear term in $Y$ we rule out based on symmetry but can be added to mimic the effect of an applied surface field.  In 4d, this $Y^2$ term is in fact the only relevant boundary deformation available to us.\footnote{%
 The old analyses \cite{PTandCPreviews,binder1983critical} typically exclude a boundary $Y^3$ term, which would be classically marginal.  
 Something interesting happens  with an interaction of the form $\mathscr{L}_{\rm bry} = g Y^\beta \delta(z)$ however.  
In order for this boundary interaction to be consistent with the profile (\ref{Phiprofile}), one must choose $\beta = n+2$ and $g = -\frac{\lambda^n}{n+2}$.  Interestingly, this correlation between bulk and boundary interactions is exactly what happens in supersymmetric theories.  The interactions are controlled by a superpotential $W = \frac{\lambda^n}{n+2} Y^{n+2}$ which leads to a potential $\frac{1}{2} \lambda^{2n} Y^{2n+2}$ along with a boundary term $-W \delta(z)$ if we are to preserve supersymmetry (see \cite{DiPietro:2015zia} and references therein).
}  
The variational principle leads us then to an additional boundary constraint on the field:
\be
- \partial_z Y + m Y = 0 \ .
\ee
For the $Y \sim z^{-1/n}$ solution, we can try to make sense of this boundary condition by introducing a small distance cut-off $z = \epsilon$ that regulates the boundary divergence as $z \to 0$.  We find a possible solution provided
\be
\frac{1}{n} \frac{1}{\epsilon} + m = 0 \ .
\ee
The equation makes sense provided $m < 0$.  Otherwise, for $m>0$ we expect $Y = 0$ to be the preferred background solution, i.e.\ Dirichlet boundary conditions.
In the literature for $Y^4$ theory with boundary, one finds special names for these cases.  The $Y = 0$ Dirichlet solution is often called ``ordinary'' while the $Y \sim z^{-1/n}$ is the  ``extraordinary'' or boundary ordered case.  At the critical point $m = 0$, one finds ``special'' or Neumann boundary conditions \cite{PTandCPreviews,binder1983critical,Diehl:1996kd}.  
Note that in the context of the renormalization group, in the IR at a small scale $\Lambda$, the dimensionless coupling $m / \Lambda$ will become effectively infinite.

If we then look for fluctuations around this ``extraordinary'' background, $Y = Y_0 + \phi$, at quadratic order, we find the Lagrangian density
\be
\label{Lexpanded}
\mathscr{L} 
&\rightarrow&  \frac{1}{2}(\partial  \phi)^2 + \frac{(1+n)(1+2n)}{2n^2 z^2}  \phi^2 \ ,
\ee
{\it et voil\`a}, a conformal mass has been generated for the fluctuations about the classical solution.
In terms of $\mu$, we have
\be
\mu = \pm \frac{3n+2}{2n} \ .
\ee
For the particular case $n=1$, one finds $\mu = \pm \frac{5}{2}$, and we could rule out the minus sign based on unitarity of the effective CFT living on the boundary, as we will discuss in more detail in the sections to come.  Computing the two point function
$\langle \phi(x) \phi(x') \rangle$, this value $\mu = \frac{5}{2}$ leads to critical exponents for the fluctuations of the order parameter $\eta_{\perp} = 3$ and $\eta_\parallel = 6$ as calculated in ref.\ \cite{Lubensky:1975zz} long ago
for the extraordinary transition.\footnote{%
For the $Y^{n+2}$ boundary interaction, for some reason the natural boundary condition picks out the minus sign, which is below the unitarity bound $\mu=-1$.
}

We emphasize in concluding that fluctuations around the field $Y$ are only one way of generating these $z$-dependent marginal couplings.  Let us first restrict to theories such that the choice of $n$ produces a classically marginal coupling in (\ref{PhiL}).
We can then take a step back and consider a more general field theory which contains additional dynamical fields that couple to
$Y$ in a classically marginal way.  The profile $Y_0$ will induce $z$-dependent classically marginal interactions for these other fields.
For example, if we had a Yukawa type coupling $Y \bar \psi \psi$ to fermions in a 4d theory, the profile for $Y_0$ would generate a conformal mass for the fermions.

\section{Dependence of the Partition Function on Marginal Parameters}
\label{sec:derivative}

In this section, we will use conformal perturbation theory to investigate the dependence of the partition function on 
marginal deformations of the form $\int \d^d x\, z^{\Delta-d} \mathcal{O}_\Delta$.
We will focus on two examples: the hemisphere and the spherical boundary.

 In a boundary CFT, the one point function of an operator is restricted by symmetry to have the form 
 \cite{McAvity:1993ue,McAvity:1995zd}
\be
\langle {\mathcal O}_\Delta \rangle = \frac{a_{\mathcal O}}{z^{\Delta}} \ ,
\ee
where $a_{\mathcal O}$ is a constant determined by the details of the theory under consideration.  Quite generally, then, we can express the derivative of the  partition function in terms of $a_{\mathcal O}$.  If we write the partition function schematically as
a Euclidean path integral
\be
Z = \int [d \phi] \exp \left(-S[\phi] - \delta S[\phi] \right) \ ,
\ee
our effective action is then $W \equiv - \log Z$.  Taking a derivative with respect to $\lambda$, we find
\be
\label{dWdlambda}
\lambda \frac{d W}{d\lambda} = \langle \delta S[\phi] \rangle \ .
\ee
In flat space, this expression suffers from a long distance divergence, but on the ball or hemisphere there are no long distance divergences, and we can regulate the short distance divergences that occur as we get close to the boundary and obtain sensible results.  On the ball, we find
\be
\langle \delta S[\phi] \rangle_{\rm ball} = \lambda \int_{r<L} \d^d x \frac{a_{\mathcal O}}{\left( \frac{L^2 - r^2}{2L} \right)^d} \ , 
\label{ballint}
\ee
while on the hemisphere we get instead
\be
\langle \delta S[\phi] \rangle_{HS^d} = \lambda \int_{HS^d} \d V(S^d) \frac{a_{\mathcal O}}{\cos^d \theta} \ .
\label{hemisphereint}
\ee
Interestingly, the result depends on $\Delta$ only through the one-point function coefficient $a_{\mathcal O}$.  
Even for free theories, $a_{\mathcal O}$ can depend on $\lambda$ in a complicated way.  
If we can establish this dependence however, then we can integrate (\ref{dWdlambda}) and
recover the full partition function.  
In the case of free fermions and free scalars, we will be able to determine $a_{\mathcal O}(\lambda)$
explicitly  and compute $W$.

Based on the Weyl scaling arguments in the introduction, we expect the integrals on the ball (\ref{ballint}) and hemisphere (\ref{hemisphereint}) to give the same answer.  (Indeed, they should also be equivalent to calculations on hyperbolic space with a spherical boundary, connecting these calculations to the AdS/CFT double trace literature \cite{Gubser:2002zh,Gubser:2002vv,Allais:2010qq}.)  Of course, short distance divergences and the necessity to add counter-terms to regulate them makes the story more nuanced.  Let us try to see this equivalence in more detail. 

Consider the BCFT defined on the hemisphere $HS^d$. 
The integral (\ref{hemisphereint}) reduces to
\be
\langle \delta S[\phi] \rangle_{HS^d} = a_{\mathcal O}  \, \lambda \, \Vol(S^{d-1})  \int_0^{\pi/2} \frac{\sin^{d-1} \theta}{\cos^d \theta}\d \theta \ ,
\label{HSintactual}
\ee
where $\Vol(S^{d-1}) = 2 \pi^{d/2} / \Gamma(\frac{d}{2})$ is the volume of a sphere of unit radius.  The integral diverges as $\theta \to \frac{\pi}{2}$ in the cases of interest, but may be evaluated using dimensional regularization.  We find
\be
\langle \delta S[\phi] \rangle_{HS^d} = a_{\mathcal O}  \, \lambda \, \pi^{\frac{d-1}{2}} \Gamma \left(\frac{1-d}{2}\right) \ .
\label{HSdimreg}
\ee
There is a divergence in dimensional regularization for odd values of $d$.  This log divergence is symptomatic of the fact that the boundary is even dimensional and may support a scale anomaly, as discussed in the introduction.

To gain further clarity on the divergence structure, we can evaluate the integral using a cut-off prescription, $0 < \theta < \frac{\pi}{2} - \delta$, instead of dimensional regularization.  In this alternate regularization scheme, we find
in dimensions two, three, and four that
\be
\label{HStwo}
\frac{1}{a_{\mathcal O} \lambda} \langle \delta S[\phi] \rangle_{HS^2} &=& \frac{2\pi}{\delta} - 2\pi + O(\delta) \ , \\
\frac{1}{a_{\mathcal O} \lambda} \langle \delta S[\phi] \rangle_{HS^3} &=&  \frac{2\pi}{ \delta^2}+ 2 \pi \log \delta + O(1) \ , \\
\frac{1}{a_{\mathcal O} \lambda} \langle \delta S[\phi] \rangle_{HS^4} &=& \frac{2\pi^2}{3 \delta^3} - \frac{5\pi^2}{3 \delta} + \frac{4\pi^2}{3} + O(\delta)\ .
\label{HSfour}
\ee
The power law divergences that appear must be in correspondence with the possible counter-terms that can be written on the boundary of the hemisphere.  As the extrinsic curvature of the boundary vanishes, we are limited to working with scalar quantities constructed from powers of the Riemann tensor.  The leading divergence scales with the volume of the boundary of the hemisphere, while subleading divergences are further suppressed by even powers of the cut-off. 
As usual these power law divergences are scheme dependent.
By power counting, 
in odd dimensions, where the boundary is even dimensional, the coefficient of the logarithm cannot be altered by
counter-terms.  In contrast in even dimensions, where the boundary is odd dimensional, the constant term
should be scheme independent and carry some physical information.
Indeed the constant and log terms match the result from dim reg (\ref{HSdimreg}).%
\footnote{We will discuss the log term that appears on the boundary in odd dimensions and explain how its coefficient can depend on marginal parameter in greater detail in \cite{toappear}, to appear soon.}

For the ball, the story is more complicated.  The boundary has extrinsic curvature, and counter terms can involve contractions of both extrinsic curvature and the Riemann curvature.   If we were to use a cut-off prescription $r<L-\epsilon$, we would find power law divergences at every order instead of every other order as we did for the hemisphere.  While the coefficient of the log in odd dimensions remains a scheme independent quantity, the constant term in even dimensions can now be shifted by counter terms.%
\footnote{In some sense, this perspective is too sophisticated for the two examples at hand.  
The integrals (\ref{ballint}) and (\ref{hemisphereint}) are identical up to a change of variables, 
$\tan \frac{\theta}{2} = \frac{r}{L}$ (see appendix \ref{app:maps}). The corresponding relation between the cut-offs $\epsilon$ and $\delta$ would allow us to map (\ref{HStwo})-(\ref{HSfour}) to results on the ball.  Even more simply, if we were to use dim reg, we would obtain (\ref{HSdimreg}) directly from (\ref{ballint}).}

Moving forward, we consider theories of free fermions and bosons with a conformal mass perturbation. We will see that as a result of bulk contact singularities, $a_{\mathcal O}$ in general contains a log divergence in even numbers of spatial dimensions, but not odd. As a consequence, the partition function will contain a log in every dimension.

\section{Free Scalar}
\label{sec:scalar}

While our principle interest is in the flat half space ${\mathbb R}^+ \times {\mathbb R}^{d-1}$, we will also compute a partition function on a hemisphere, and it is useful to take a more general curved space point of view.
We start with the action for a conformally coupled scalar field $\phi(x)$ on an arbitrary curved space-time $M$ with boundary $\partial M$:
\be
S_0 = \frac{1}{2} \int_M \d^d x \sqrt{g} [ (\partial \phi)^2 + \beta R \phi^2 ] + \beta \int_{\partial M} \d^{d-1} x \sqrt{\gamma} K \phi^2 \ ,
\ee
where $\sqrt{g}$ is the square root of the determinant of the metric on $M$, $\sqrt{\gamma}$ the same for $\partial M$, 
$R$ is the Ricci scalar curvature, and $K$ is the trace of the extrinsic curvature on the boundary.
The conformal value of the coupling is $\beta = \frac{d-2}{ 4 (d-1)}$.  The boundary term is required to preserve Weyl invariance.  
We perturb this action by adding a ``mass term'' 
\be
\delta S = \frac{1}{2} \left(\mu^2 - \frac{1}{4}\right) \int \d^d x \sqrt{g} \, Y(x)^2 \phi(x)^2 \ ,
\ee
that maintains Weyl invariance because we insist that the conformal compensator $Y(x)$ transform as $Y \to e^{-\sigma} Y$ under rescalings of the metric, $g_{\mu\nu} \to e^{2 \sigma} g_{\mu\nu}$. 
The stress-energy tensor is 
\begin{align} \label{}
T_{\mu\nu} = (\partial_\mu \phi)( \partial_\nu \phi) -  \frac{\eta_{\mu\nu}}{2} \left( (\partial \phi)^2 + \left(\mu^2 - \frac{1}{4}\right) Y^2 \phi^2 \right) - \beta(\partial_\mu \partial_\nu - \eta_{\mu\nu} \partial^2 ) \phi^2 \ ,
\end{align}
and
\begin{align} \label{}
\mathcal{O} = - \left( \mu^2 - \frac{1}{4} \right) Y\phi^2 \ . 
\end{align} 
One can verify using the equations of motion that these expressions satisfy the Ward identities (\ref{Ward_identity}). To preserve conformal symmetry we choose in flat space $Y = z^{-1}$ as explained in \ref{sec:symmetries}.

On the half space, we will compute the two-point correlation function $\langle \phi(x) \phi(x') \rangle$ from which we will be able to extract the coefficient of the one-point function $a_{\phi^2}$ by an appropriate regularization, allowing us to compute the effective action $W(\mu)$ by integrating (\ref{dWdlambda}).  
On the hemisphere, we will compute the partition function directly by diagonalizing the equations of motion.

\subsection{Two Point Functions on the Half Space}

First some preliminary definitions.  We use a coordinate system $x = (z,{\bf x})$ where $z>0$ is the distance from the boundary and ${\bf x}$ are tangential directions.
It is also convenient to define a pair of related cross ratios
\be \label{inv_ratios}
v^2 \equiv \frac{({\bf x}-{\bf x}')^2 + (z-z')^2}{({\bf x}-{\bf x}')^2 + (z+z')^2} \ , \; \; \; \xi \equiv \frac{v^2}{1-v^2} \ .
\ee
There are two interesting limits for $v$, the boundary limit $v \to 1$, and the coincident limit $v \to 0$.

From the constraints of the residual boundary conformal group $SO(d,1)$, we know that the two point function for $\phi$ must have the general form \cite{McAvity:1993ue,McAvity:1995zd}
\be
\langle \phi(x) \phi(x') \rangle = \frac{g(v)}{|x-x'|^{d-2}} \ .
\ee
Because the action is quadratic, the two-point function can be calculated classically from the equation of motion:
\be
\left[ \Box_x - \left( \mu^2 - \frac{1}{4} \right) \frac{1}{z^2} \right] \langle \phi(x) \phi(x') \rangle = \delta(x-x') \ .
\ee
This equation of motion implies that $g(v)$ satisfies the following ordinary differential equation,
\be
(1-v^2)^2 v g''(v) - (d-3) (1-v^2)^2  g'(v) - \left( \mu^2 - \frac{1}{4} \right) v g(v) = 0 \ ,
\ee
which can be solved in terms of hypergeometric functions.
The integration constants can be determined from the behavior of the Green's function in the boundary and coincident limits.
In the coincident $v \to 0$ limit, we expect the effect of the conformal mass to be negligible and we should recover the familiar result
\be
 \langle \phi(x) \phi(x') \rangle \sim \frac{\kappa}{|x-x'|^{d-2}} \ ,
\ee
where a conventional normalization is that $\kappa^{-1} =(d-2) \Vol(S^{d-1})$.  
In the boundary $v\to 1$ limit, in general one has a linear combination of two possible behaviors 
$g(v) \sim (1-v)^{\frac{1}{2} \pm \mu}$.  
We demand that $g(v) \sim (1-v)^{\frac{1}{2}+ \mu}$ and that the other behavior cancels out.  
The conformal mass term vanishes when $\mu = \pm \frac{1}{2}$.  From the boundary behavior, 
we interpret $\mu = \frac{1}{2}$ as Dirichlet boundary conditions and $\mu = -\frac{1}{2}$ as Neumann.  

Given the boundary conditions, we find the unique solution
\be
\label{gvfinal}
g(v) =  \kappa 
 \frac{\Gamma \left( \frac{1}{2} +\mu\right) \Gamma \left( \frac{d-1}{2} +\mu\right) }{\Gamma \left( \frac{d}{2}-1 \right) \Gamma(1+2\mu)}
 \xi^{-\frac{1}{2}-\mu} {}_2 F_1 \left( \frac{1}{2}+\mu, \frac{d-1}{2}+\mu, 1+2\mu, -\frac{1}{\xi} \right) \ .
\ee
The other boundary behavior $g(v) \sim (1-v)^{\frac{1}{2} - \mu}$ occurs after the replacement $\mu \to -\mu$.
As the action depends only on $\mu^2$, we adopt the scheme of classifying the scalar 
not by the $\mu^2$ parameter in the action but by the near boundary fall-off $\mu$. 

One aspect of this analysis that so far remains obscure is the allowed range of $\mu$.  One way establish this range is to compute the Green's function instead by diagonalizing the operator
\be
-\Box + \left( \mu^2 - \frac{1}{4} \right) \frac{1}{z^2} \ .
\ee
If we look for eigenmodes of the form $\varphi(x) = e^{i k \cdot {\bf x}} f(z)$, we find that $f(z)$ has a Bessel function form
\be
f(z) = \sqrt{z} J_\mu(E z) \ ,
\ee
with eigenvalue $E^2 + k^2$. 
These modes are plane wave normalizable far from the boundary.  Additionally, they are normalizable at the boundary provided $\mu > -1$.  

We could have obtained (\ref{gvfinal}) more easily by copying the corresponding AdS/CFT result eq.\ (2.5) of \cite{DHoker:1999mqo}, the bulk-to-bulk propagator, with the replacement
\be
\label{muDelta}
\mu = \Delta - \frac{d-1}{2} \ .
\ee
It is straightforward to relate $g(v)$ to the two-point function for a scalar field in hyperbolic space (or equivalently $AdS_d$ in the Lorentzian setting)
\cite{Buhl-Mortensen:2016jqo}.  
If $\phi(x)$ is a field in ${\mathbb R}^+ \times {\mathbb R}^{d-1}$ and $\tilde \phi(x)$ is a field in $H^d$, 
based on the Weyl scaling arguments in the introduction, we expect $\tilde \phi(x) = (z/L)^{(d-2)/2} \phi(x)$.  
Thus the two-point functions will be related via
\be
\langle \tilde \phi(x) \tilde \phi(x') \rangle = \left(\frac{z z'}{L^2} \right)^{(d-2)/2} \langle \phi(x) \phi(x') \rangle = \left(\frac{1}{4\xi L^2}  \right)^{(d-2)/2} g(v) \ ,
\ee
where we have succeeded in writing the correlation function completely in terms of $SO(d,1)$ invariant quantities of \eqref{inv_ratios}, $v$ or equivalently $\xi$. The unitarity bound $\Delta > \frac{d-3}{2}$ in the AdS/CFT literature corresponds to our bound $\mu > -1$ that we arrived at through normalizability considerations.

The expression (\ref{gvfinal}) has an interesting consequence for the boundary operator product expansion, one that could have been anticipated given the relation to AdS/CFT \cite{Carmi:2018qzm}.  In a boundary CFT, we expect to be able to decompose any bulk operator into a sum over boundary operators.  This decomposition allows in turn an expression of the two-point function of two bulk operators as a sum over two-point functions of boundary operators and their descendants.  The sum over a boundary primary operator and its descendants can be re-organized into a boundary conformal block.  The total sum can then be condensed into a sum over the conformal blocks for the boundary primaries.  For bulk operators ${\mathcal O}$ of dimension $\eta$, the generic form of this expansion is (see e.g.\ \cite{Herzog:2017xha} or \cite{Liendo:2012hy})
\be
\langle {\mathcal O}(x) {\mathcal O}(x') \rangle = \xi^\eta \left[ a_{{\mathcal O}}^2 + \sum_{\Delta \neq 0} \mu_\Delta^2 G_{\rm bry} (\Delta, v) \right]
\ee
where the boundary conformal blocks are
\be
G_{\rm bry}(\Delta, v) = \xi^{-\Delta} {}_2 F_1 \left( \Delta, 1 - \frac{d}{2} + \Delta, 2-d + 2 \Delta, -\frac{1}{\xi} \right) \ .
\ee
In our case ${\mathcal O}(x) = \phi(x)$ and the one-point function for a single $\phi(x)$ vanishes, $a_{\phi} = 0$.
From the solution (\ref{gvfinal}) for $\langle \phi(x) \phi(x') \rangle$,  we see that the sum over boundary conformal blocks reduces to a single block, namely for a boundary operator of dimension $\Delta = \mu + \frac{d-1}{2}$.  

This boundary OPE reveals the nature of the effective theory living on the boundary.  If we were just interested in correlation functions restricted to the boundary and wanted to forget about the bulk physics, we learn that we have a single generalized free field $\varphi({\bf x})$ with scaling dimension $\Delta$.  Higher correlation functions follow from Wick's theorem and the result for the two point function, that $\langle \varphi({\bf x}) \varphi({\bf 0}) \rangle \sim |{\bf x}|^{-2 \Delta}$.  In the limit $\Delta \to \frac{d-3}{2}$, we reach the unitarity bound; the boundary field theory should become that of an actual free scalar field.

Given that the full boundary theory can be reconstructed from the Green's function $|{\bf x}|^{-2 \Delta}$, we should be able to compute the partition function from the Green's function.  One way of defining a fractional Laplace operator $(-\Box)^{\alpha}$ is as the inverse of the Green's function $|x|^{-2\Delta}$ with $\alpha = \frac{d}{2} - \Delta$  \cite{Kwasnicki}.  
The partition function is then by definition the determinant of such a fractional differential operator.
When we calculate the full partition function below, we will be able to isolate a boundary contribution to it, which we can tentatively identify as the partition function of such a fractional Laplacian.

As a first pass toward the computation of $W(\lambda)$, we will extract $a_{\phi^2}$ by taking the coincident limit of the 
two-point function $\langle \phi(x) \phi(x') \rangle$.  
By a hypergeometric identity, the result (\ref{gvfinal}) can be written as
\be
g(v) &=& \kappa (1-v^2)^{ \frac{1}{2}-\mu} {}_2 F_1 \left(  \frac{1}{2} -\mu,  \frac{3-d}{2}-\mu, 2 - \frac{d}{2}, v^2 \right)
\nonumber \\
&& + c v^{d-2} (1-v^2)^{ \frac{1}{2}-\mu} {}_2 F_1 \left( \frac{1}{2}-\mu , \frac{d-1}{2}- \mu, \frac{d}{2}, v^2 \right) \ .
\label{gvtwoterm}
\ee
where 
\be
c = \kappa  \frac{\Gamma \left( 1 - \frac{d}{2} \right) \Gamma \left( \frac{d-1}{2} + \mu\right)}
{ \Gamma \left( \frac{d}{2} -1\right)\Gamma \left( \frac{3-d}{2}+ \mu \right) }\ .
\ee
We use point splitting and subtract the divergences associated with the coincident limit.  There are several such divergences, associated with the first term in (\ref{gvtwoterm}).  After subtraction, the leading and in general finite contribution will come from the second term in (\ref{gvtwoterm}).  We find then that
\be
\label{aphi2}
\langle \phi^2(x) \rangle = \frac{a_{\phi^2}}{z^{d-2}} \; \; \mbox{   where   } \; \; a_{\phi^2} =  \kappa \frac{\Gamma \left( 1 - \frac{d}{2} \right) \Gamma \left( \frac{d-1}{2} + \mu\right)}
{2^{d-2} \Gamma \left( \frac{d}{2} -1\right) \Gamma \left( \frac{3-d}{2}+ \mu \right) } \ .
\ee
Notably, there is a divergence in $a_{\phi^2}$ in even dimensional spaces, associated 
with the pole in $\Gamma\left(1-\frac{d}{2}\right)$.  
This divergence means that the derivative of the effective action $\partial_{\lambda} W$ will be log divergent in general dimension and not just in odd dimension.  In odd dimension, the logarithmic divergence can be attributed to a boundary divergence in the integral over the one-point function.  In even dimension, the log divergence comes from the one-point function itself and its normalization $a_{\phi^2}$. 

These log divergences in $a_{\phi^2}$ are absent if we first take $\mu \to \pm \frac{1}{2}$, i.e.\ the Dirichlet and Neumann cases.  In these cases where the conformal mass vanishes, we find the familiar results for the one-point function, that
\be
\langle \phi(x)^2 \rangle  \to \mp \frac{\kappa}{(2 z)^{d-2}} \ ,
\ee
as we could have obtained from the method of images, for example.

Assembling (\ref{dWdlambda}), (\ref{HSintactual}), and (\ref{aphi2}), we can write down an integral expression for the partition function.  Re-expressing $\lambda$ in terms of $\mu$, we obtain in dimensional regularization
\be
\label{Wdiffint}
W(\mu) - W(0) = \Gamma(1-d)  \int_0^\mu \frac{\Gamma \left(\frac{d-1}{2} + \mu' \right)}{\Gamma\left(\frac{3-d}{2} + \mu' \right) } \mu'  \d \mu' \ .
\ee 
The regularization scheme will suppress the power law divergences we saw in the previous section, but allows for logarithmic ones.  Indeed, as claimed, we see a log divergence in every positive integer dimension $d \geq 1$ associated with the pole in the 
$\Gamma(1-d)$ function.  The integrand is a degree $d-1$ polynomial in $\mu'$ for general integer $d$ and is straightforward to evaluate.

This result (\ref{Wdiffint}) is equivalent by Weyl scaling to previous calculations of the partition function of a massive scalar in AdS (see e.g.\ \cite{Kamela:1998mb,Gubser:2002zh}). Nevertheless, we believe that our novel context -- of marginal deformations in boundary CFT -- is different enough to have made a self-contained derivation of (\ref{Wdiffint}) worthwhile.  In fact, we will now present a second derivation, computing the partition function on the hemisphere directly.
The formula (\ref{Wdiffint}) has many notable features, some of which we review and explore below and one of which is that 
expanding out the integrand near an integer dimension, $d = n-\epsilon$, 
the integral can be re-expressed in terms of Barnes' multiple gamma functions \cite{Kamela:1998mb,Basar:2009rp,Aros:2010ng}.  
A virtue of the hemisphere partition function calculation is that the sum over eigenvalues gives precisely the definition of the multiple gamma function, without the need to manipulate or massage the expression (\ref{Wdiffint}).

\subsection{The Partition Function on the Hemisphere}
\label{sec:scalarHS}

 If we work on a hemisphere of unit radius, then the Ricci scalar curvature is  $R = (d-1) d$ and the trace of the extrinsic curvature vanishes, $K=0$.  The action becomes
\be
S_0 = \frac{1}{2} \int_{HS^d} \d V(S^d) \left[ (\partial \phi)^2 + \frac{d(d-2)}{4} \phi^2 \right] \ , 
\ee
to which we add the marginal term
\be
\delta S = \frac{\mu^2-\frac{1}{4}}{2} \int_{HS^d} \d V(S^d) \frac{\phi^2}{\cos^2 \theta} \ ,
\ee
where we write the metric on the hemisphere in terms of the line element on a sphere of one dimension less, $\d \Omega_{d-1}^2$:
\be
\label{spheremetric}
\d s^2 = \d \theta^2 + \sin^2 \theta \, \d \Omega_{d-1}^2 \ .
\ee

The theory is quadratic and the partition function is a Gaussian integral.  Writing the equation of motion as ${\mathcal D} \phi = 0$, the effective action will be $W = \frac{1}{2} \log \det {\mathcal D} + c$ up to some overall normalization of the path integral corresponding to $c$.  We need then to 
determine the eigenvalues of ${\mathcal D}$, 
\be
{\mathcal D} = -\frac{1}{\sin^{d-1} \theta} \partial_\theta (\sin^{d-1} \theta  \partial_\theta) - \frac{1}{\sin^2 \theta} \Box_{d-1} + \frac{d(d-2)}{4}  + \frac{\mu^2 - \frac{1}{4}}{\cos^2 \theta} \ ,
\ee
where $\Box_{d-1}$ is the Laplacian on a $S^{d-1}$. 

 Let us first review some well known results for the eigenvalues and eigenfunctions of the Laplacian on an $S^d$. 
The eigenfunctions are  generalized spherical harmonics $Y_{\vec \ell} (\vec \theta)$, where $\vec \ell = (\ell_1, \ldots, \ell_{d})$ and $\vec \theta = (\theta_1, \ldots, \theta_{d})$ where $\theta_d = \theta$.  The allowed values of $\vec \ell$ are integers such that $|\ell_1| \leq \ell_2 \leq \cdots \leq \ell_d$. To simplify notation, we will set $\ell = \ell_d$ and hopefully without confusion $m = \ell_{d-1}$. Furthermore $\Box_{d} Y_{\vec \ell} (\vec \theta) = - \ell (\ell +d-1)$.
These eigenvalues have degeneracy
\be \label{scalar_deg}
\dim(\ell) = { d+\ell \choose d} - {d+\ell-2 \choose d} \ .
\ee

In the case of a hemisphere without a conformal mass, we can identify eigenfunctions on the full sphere which have the appropriate boundary conditions to be eigenfunctions on the hemisphere as well.  The spherical harmonics $Y_{\vec \ell} (\vec \theta)$ where $\ell - m \in 2 {\mathbb Z}+1$ will be odd functions about the equator of the sphere.  Hence, they satisfy Dirichlet boundary conditions.  The spherical harmonics with $\ell - m \in 2 {\mathbb Z}$ are even functions about the equator and satisfy Neumann boundary conditions.  From this parity restriction and the corresponding eigenvalue degeneracies on an $S^{d-1}$, 
one finds degeneracy for modes on the hemisphere.
\be \label{scalar_N_D}
\dim_N(\ell) = {d-1+\ell \choose d-1} \ , \; \; \; \dim_D(\ell) = {d-2+\ell \choose d-1}
\ee
along with an identity $\dim(\ell) = \dim_N(\ell) + \dim_D(\ell)$.  Here $N$ stands for Neumann and $D$ for Dirichlet.

Moving now to the case of a nonzero conformal mass, if we let $Y_{\vec \ell} (\vec \theta)$ be an eigenfunction on $S^{d-1}$, instead of $S^d$ as we did above, we can look for an eigenfunction on the hemisphere of the form $\phi(x) = g(y) Y_{\vec \ell} (\vec \theta)$ where $y = \cos^2 \theta$.  We find two well behaved solutions.
One set of candidate eigenfunctions of ${\mathcal D}$ which are regular on the north pole $y=1$ are
\be
\label{gNP}
g_{\ell,m}^{\rm NP} = (1-y)^{\frac{m}{2}} y^{\frac{\mu}{2}+ \frac{1}{4}}
{}_2 F_1 \left( -\frac{\ell-m}{2}, \frac{d+\ell+m}{2}+\mu, \frac{d}{2} + m, 1-y \right) \ .
\ee
Another set which have the $y^{\frac{\mu}{2} + \frac{1}{4}}$ behavior at the equator $y=0$ are
\be
\label{gEQ}
g_{\ell,m}^{\rm EQ} = (1-y)^{\frac{2-d-m}{2}} y^{\frac{\mu}{2} + \frac{1}{4}} 
{}_2 F_1 \left( 1 - \frac{d+\ell+m}{2}, \frac{\ell-m}{2} + \mu+1, 1+\mu, y\right) \ .
\ee
As we found in flat space $\mu = - \frac{1}{2}$ corresponds to Neumann boundary conditions and $\mu = \frac{1}{2}$ to Dirichlet.
These two candidates (\ref{gNP}) and (\ref{gEQ}) are equivalent when $\ell-m$ is a non-negative even integer:
\be
\frac{g_{\ell,m}^{\rm NP}(y)}{g_{\ell,m}^{\rm EQ}(y)} = \frac{i^{\ell-m} \Gamma \left( \frac{d}{2} + m \right) \Gamma \left( 1 + \frac{\ell-m}{2} + \mu \right)}{\Gamma\left( \frac{d+\ell+m}{2}\right) \Gamma(1+\mu)} \ .
\ee
The corresponding eigenvalues are
\be
E(\mu, \ell)  = \left( \ell + \mu + \frac{d-1}{2} \right) \left( \ell + \mu + \frac{d+1}{2} \right) \ .
\ee
The conformal mass thus introduces a simple overall shift in the eigenvalues.
The parity restriction on $\ell-m$ means that the degeneracy of these eigenvalues is the same $\dim_N(\ell)$ function as for the Neumann boundary condition case in the absence of a conformal mass.  
We get the correct degeneracy counting in the Dirichlet case as well.  Shifting $\mu \to \mu+1$ is equivalent to shifting $\ell \to \ell+1$ in $E(\mu, \ell)$.  We then have the identity that $\dim_N(\ell) = \dim_D(\ell+1)$.  

The basic object that we need to compute is the divergent quantity
\be
\label{Wsumform}
W(\mu) = \frac{1}{2} \sum_{\ell=0}^\infty \dim_N(\ell) \log E(\mu, \ell) + c\ .
\ee
This object is naturally written as a Barnes' multiple gamma function \cite{Barnes}.
The Barnes' multiple zeta function is defined as a sum over integers
\be
\zeta_d(s,a) \equiv \sum_{m_1, \ldots, m_d \geq 0} \frac{1}{(m_1 + \cdots + m_d + a)^s} \; , \; \; ({\rm Re}(s) > d) \ ,
\ee
which generalizes the Hurwitz zeta function $\zeta_1(s,a) = \zeta(s,a)$.  
This multiple sum can be reduced to a single sum
\be
\zeta_d(s,a) = \sum_{\ell=0}^\infty {\ell+d-1 \choose d-1} \frac{1}{(\ell+a)^s} \ ,
\ee
where the binomial coefficient can be identified as the degeneracy $\dim_N(\ell)$. 
The multiple gamma function we need is 
\be
\log \Gamma_{d}(a) \equiv \lim_{s \to 0} \partial_s \zeta_d(s,a) \ .
\ee
From these definitions, we find that
\be
W(\mu) =  -\frac{1}{2} \log \Gamma_d\left( \mu + \frac{d-1}{2} \right) - \frac{1}{2} \log \Gamma_d \left( \mu + \frac{d+1}{2} \right) + c \ ,
\ee 
providing another route to the results \cite{Kamela:1998mb,Basar:2009rp,Aros:2010ng} obtained by direct evaluation of (\ref{Wdiffint}).

\subsection{Evaluation of the Partition Function}

Although elegant, we do not need the Barnes' interpretation of (\ref{Wsumform}) to proceed.
We can regularize some of the divergences by considering the difference
\be
W(\mu) - W(0) = \frac{1}{2} \sum_{\ell=0}^\infty \dim_N(\ell) \log \frac{E(\mu, \ell)}{E(0, \ell)} \ .
\ee
The logarithmic divergences can be isolated approximating the sum as an integral up to some cut-off angular momentum $\ell_{\rm max}$:
\be
\frac{1}{2} \int_0^{\ell_{\rm max}} {d-1+\ell \choose d-1} \log 
\left[ \left( 1+ \frac{\mu}{\ell + \frac{d-1}{2}} \right) \left( 1+ \frac{\mu}{\ell + \frac{d+1}{2} } \right) \right] \d \ell \ .
\ee
We present some tricks for dealing with this integral in appendix \ref{app:tricks}.  We find at the end of the day that the log divergence 
can be isolated,
\be
\label{logdiv}
W(\mu) - W(0) \sim - \log \ell_{\rm max} \int_0^\mu  \frac{\Gamma \left( \frac{d-1}{2} - \mu' \right)}{\Gamma(d) \Gamma \left( \frac{3-d}{2} - \mu' \right)} \mu' \d \mu'  \ .
\ee
The log divergence matches that of (\ref{Wdiffint}) if we interpret the $\log \ell_{\rm max}$ as $\frac{1}{\epsilon}$ in dimensional regularization.

We make two remarks.  First, the integrand of (\ref{logdiv}) has the following expansion near $\mu = \pm \frac{1}{2}$:
\be
\frac{\mu \Gamma \left( \frac{d-1}{2} - \mu\right)}{\Gamma(d) \Gamma \left( \frac{3-d}{2} - \mu \right)} 
= -\frac{2\Gamma \left(\frac{d}{2}\right)}{\Gamma \left( 2- \frac{d}{2}\right) \Gamma(d)} + O\left(\mu \pm \frac{1}{2} \right) \ .
\ee
The leading term in this expansion vanishes in all even dimensions greater than two.  Consequently, the $\mu$ derivative of the log contribution to $W(\mu) - W(0)$ vanishes at leading order close to $\mu = \pm \frac{1}{2}$.  

We remark further that
\be
\label{deltaW}
W(\mu) - W(-\mu) 
\ee
does not have a log contribution in even dimensions.  
The integrand of (\ref{logdiv}) has a definite parity under $\mu \to -\mu$ that depends on the dimension.
In even dimensions, the integrand of (\ref{logdiv}) contains only odd powers of $\mu$ and is thus an odd function of $\mu$.  

The quantity $W(\mu) - W(-\mu)$ has been computed in the AdS/CFT literature in the context of double-trace deformations \cite{Gubser:2002vv}.
In our language, the idea is conceptually very similar, namely to add a positive mass term $\phi^2$ on the boundary that causes a renormalization group flow from the theory with ``Neumann'' boundary conditions $-\mu < 0$ to the theory with ``Dirichlet'' boundary conditions $\mu>0$.  For the $-\mu$ theory, the boundary operator $\phi^2$ has dimension $d-1-2 \mu$, and hence we expect it to be relevant in the language of the boundary renormalization group flow.  
For the AdS/CFT interpretation, in the odd dimensional cases where $W(\mu) - W(-\mu)$ has a log divergence, the coefficient of the log was interpreted as an RG flow induced change in the central charge of the strongly interacting CFT living on the boundary. The quantity can be verified to behave monotonically under RG flow;
for odd integer $d$, the integrand is a polynomial in $\mu$ with no zeroes in the interval $0<\mu<1$, where $\mu=1$ means the $-\mu$ theory has reached the unitarity bound for the boundary scalar field, $\Delta_- = -\mu + \frac{d-1}{2}$. We mention in passing the result in 3d, where $W(-\mu) - W(\mu) = \frac{1}{3} \mu^3 \log \ell_{\rm max}$, which in the limit $\mu \to 1$ gives the anomaly contribution to the effective action for a real scalar field in 2d.  The case $\mu = \frac{1}{2}$ in contrast can be matched with the $b$-anomaly boundary contribution to the trace anomaly for a real scalar field in 3d \cite{Jensen:2015swa}.

In the cases where the log vanishes, we may try to extract the constant contribution $W(\mu) - W(-\mu)$, using zeta function regularization.  
This difference (\ref{deltaW}) with $\mu=\frac{1}{2}$ has appeared in the literature in the $d=4$ case in another guise, as a way of computing a quantity that decreases under boundary renormalization group flow \cite{Gaiotto:2014gha}, 
\be
\label{F}
\frac{|Z_{HS^4}|^2}{Z_{S^4}} = \exp \left( a_3 \frac{L^3}{\epsilon^3} + a_1 \frac{L}{\epsilon} - 2 F_\partial \right) \ .
\ee
The ratio is designed to cancel any divergences which have a four dimensional origin, while the hemisphere guarantees that no additional logarithms associated with the extrinsic curvature of the boundary will appear.  The constant term $F_\partial$ is then the candidate monotonic function under RG flow.

If we were to take the particular case of a free scalar field with Neumann boundary conditions on the $HS^4$, then we can make the replacement $Z_{HS^4} = e^{-W\left(-\frac{1}{2}\right)}$.  The partition function on the whole sphere, on the other hand, is a sum of the Dirichlet and Neumann cases, as we saw explicitly when we diagonalized the Laplacian.  We find then that
\be
2 \log Z_{HS^4} - \log Z_{S^4} = W\left(\frac{1}{2}\right) - W\left(-\frac{1}{2}\right) \ .
\ee
If we were to consider Dirichlet boundary conditions instead, we simply change the sign of this expression.
For general $\mu$ however, the equator is singular, and it's not clear how to make sense of $Z_{S^4}$.  The natural generalization seems to be instead the constant term in the difference (\ref{deltaW}).

We will study the $d=4$ case in detail, using zeta function regularization.  
We define the auxiliary functions
\be
f_d(s) =  \frac{1}{2} \sum_{\ell=0}^\infty {3 + \ell \choose 3} \left( \frac{1}{\left( \ell+ \mu + \frac{3}{2} \right)^s} + \frac{1}{\left(\ell+\mu + \frac{5}{2} \right)^s } \right) \ .
\ee
The regularized log of the partition function is
\be
W(\mu)  =  - \lim_{s \to 0} \partial_s f_d(s)  + c \ .
\ee
We then re-express $f_d(s)$ in terms of the Hurwitz zeta function
\be
\zeta(s,a) \equiv \sum_{n=0}^\infty \frac{1}{(n+a)^s} \ .
\ee
In particular, we have
\be
\sum_{\ell=0}^\infty \frac{(\ell+1)(\ell+2)(\ell+3)}{6(\ell+a)^s} &=& \frac{1}{6} \Bigl(
(6- 11 a+ 6 a^2 - a^3) \zeta(s,a) + (11 - 12 a+ 3 a^2) \zeta(s-1,a) \nonumber \\
&&
+ 3 (2-a) \zeta(s-2, a) + \zeta(s-3,a)
\Bigr) \ .
\ee
The difference in the case $\mu = \frac{1}{2}$ can be computed exactly, yielding
\be
W\left(-\frac{1}{2}\right) - W\left(\frac{1}{2}\right) = \frac{\zeta(3)}{8 \pi^2}
\ee
in agreement with \cite{Gaiotto:2014gha}.  If we think of this difference in terms of the double trace deformation
and a flow from Neumann boundary conditions in the UV to Dirichlet boundary conditions in the IR, we have verified that indeed
the difference is positive, that $F_\partial$ decreases under RG flow.  Another special case is $\mu=1$ where the difference achieves
a maximum:
\be
W(-1) - W(1) = \frac{1}{8} \left( \log(2) - \frac{3 \zeta(3)}{2 \pi^2} \right)  > 0\ ,
\ee
which is the free energy of a real scalar field on an $S^3$ \cite{Klebanov:2011gs}.
The value $\mu=1$ is special because here the scaling dimension of the ``Neumann'' boundary operator $\Delta_- = -\mu + \frac{3}{2}$ reaches the 3d unitarity bound for scalars, $\Delta_- = \frac{1}{2}$.  

\begin{figure}
\begin{center}
\includegraphics[width=3in]{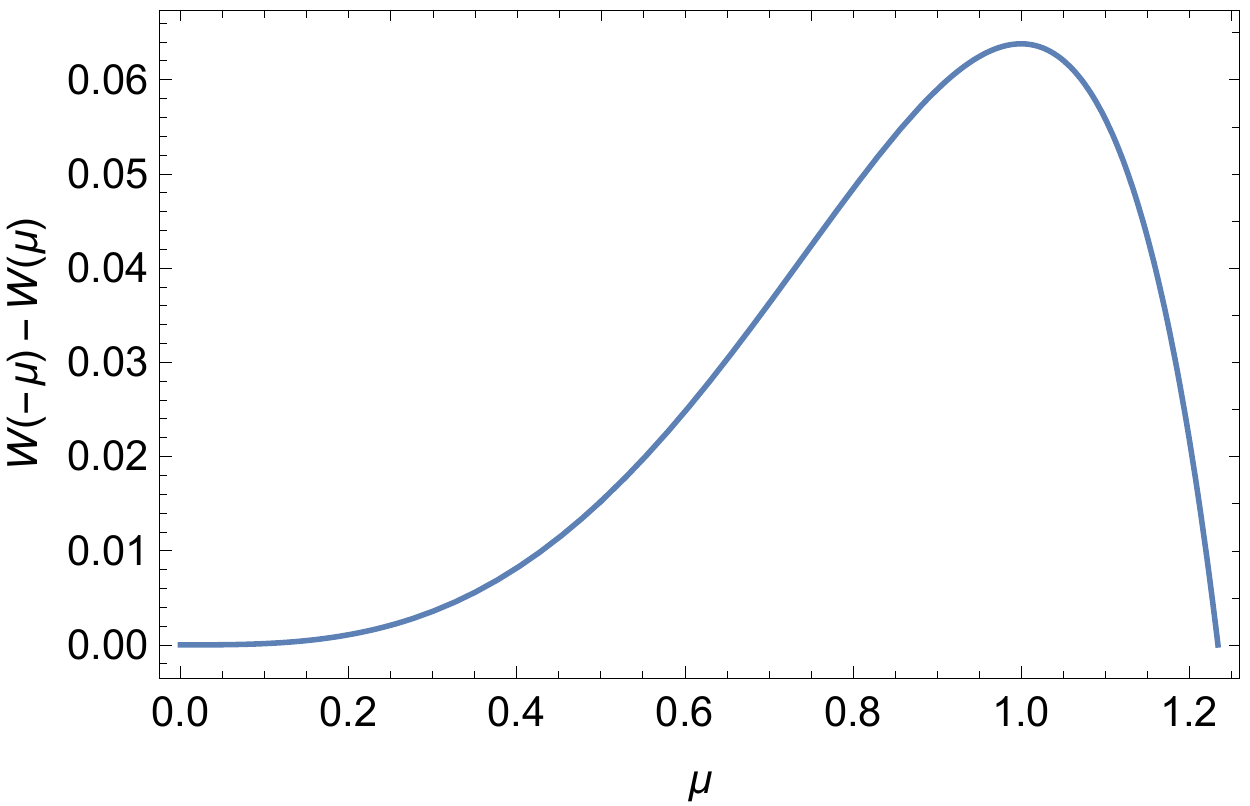}
\includegraphics[width=3in]{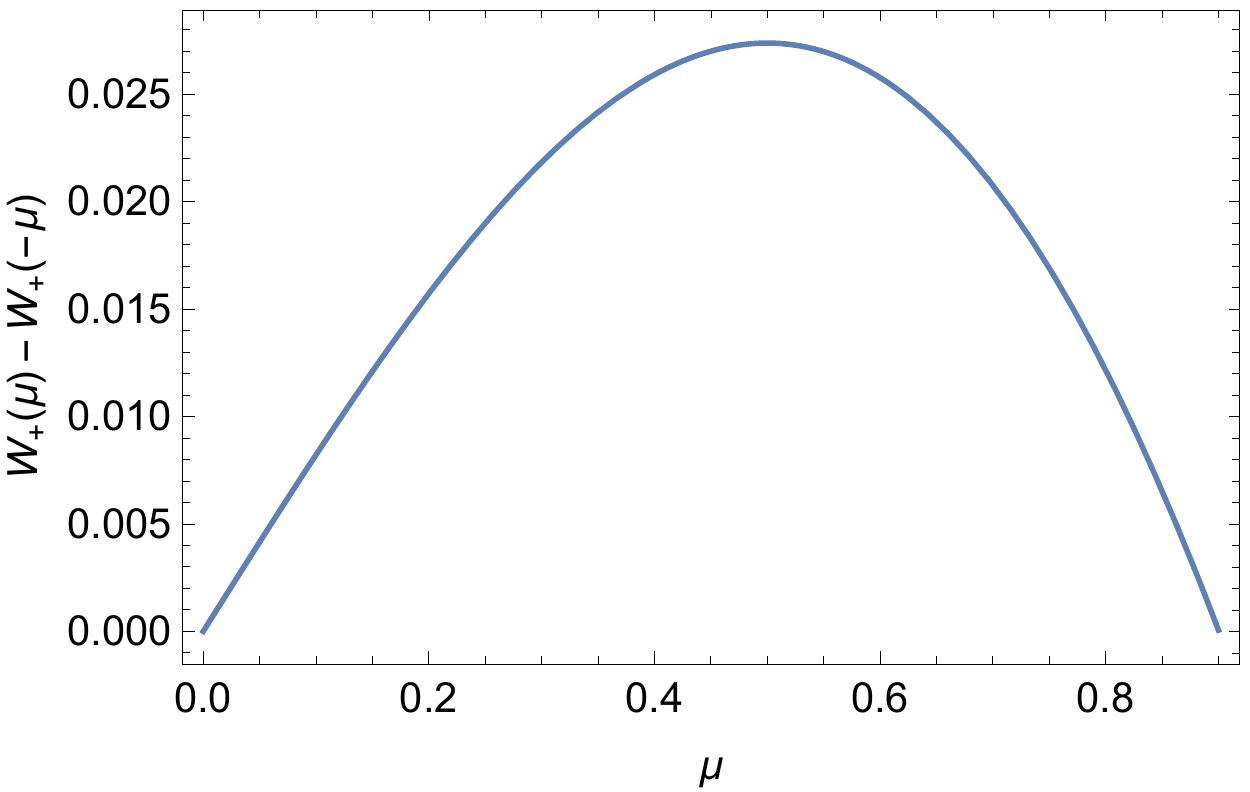}
\end{center}
\caption{(Left) The difference $W(-\mu) - W(\mu)$ as a function of $\mu$ in 4d for a free scalar.   The curve crosses the $\mu$ axis at approximately $1.23423$. (Right) The difference $W_+(\mu) - W_+(-\mu)$ as a function of $\mu$ in 4d for a free fermion.   The curve crosses the $\mu$ axis at approximately $0.90$
\label{fig:constantplot}}
\end{figure}

More generally, we obtain a messy expression for the constant term in the difference (\ref{deltaW}) involving special values of the generalized Rieman zeta function and its derivatives.  We present the result as a plot (see figure \ref{fig:constantplot}).
The curve is manifestly positive in the region $0 < \mu < 1.23423$, becomes negative for $\mu > 1.23423$, and then diverges in the limit $\mu \to \frac{3}{2}$.  

We could perform a similar computation in other dimensions, but they can be found in various guises in the literature, and we will spare the reader.

\section{Free Fermion}
\label{sec:fermion}

Let us now consider a fermion with a conformal mass.
We follow the same pattern as we did for the scalar. We start with the half space ${\mathbb R}^+ \times {\mathbb R}^{d-1}$ and
then compute the partition function directly by diagonalizing the Dirac operator on the hemisphere in the presence of the conformal mass.  For technical reasons, our direct calculation of the hemisphere partition function is restricted to the even dimensional case and a mass term that involves the chirality matrix $\gamma$.  Our Clifford algebra conventions are that
\be
\{ \gamma_a, \gamma_b \} = 2 \delta_{ab} \ .
\ee
 We define $\gamma$ so that it squares to 1 in any even dimension.

\subsection{The Half Space}

We would like to construct the Green's function for the Dirac operator
\be
\label{flatDirac}
\left(\gamma^\nu \partial_\nu + e^{i \eta \gamma} \frac{\mu}{z} \right) \psi = 0 \ ,
\ee
where $\eta$ is a real number, $\gamma$ is the gamma five chirality matrix, and $\mu$ parametrizes the strength of the conformal mass deformation. 
The factor $e^{i \eta \gamma}$ allows for a linear combination of the two types of conformal masses discussed in the introduction.  
 (In the odd dimensional cases, we take $\eta = 0$.) The coordinate $z$ is the distance from the boundary, as usual. 

Our strategy will be to use the Green's function we constructed earlier for the scalar.  Note first that
\be \label{lap_fer}
\lefteqn{\left(\gamma^\nu \partial_\nu + e^{i \eta \gamma} \frac{\mu}{z} \right) \left(\gamma^\rho \partial_\rho - e^{-i \eta \gamma} \frac{\mu}{z} \right)
 = \Box + \gamma^z e^{-i \eta \gamma} \frac{\mu}{z^2} - \frac{\mu^2}{z^2}} \nonumber \\
&=& \left( \Box - \frac{\mu(\mu-1)}{z^2} \right) \Pi_+ + \left( \Box - \frac{\mu(\mu+1)}{z^2} \right) \Pi_- \ ,
\ee
with now $\Pi_\pm \equiv \frac{1}{2} (1 \pm \gamma^z e^{-i \eta \gamma} )$.
We have the Green's function for the scalar wave equation, 
\[
\left( \Box - \frac{\mu^2-\frac{1}{4}}{z^2} \right) G_\mu(x,x') = \delta(x-x') \ .
\]
It follows that a Green's function for the fermion is
\be
G^F(x,x') = \left( \gamma^\nu \partial_\nu - e^{-i \eta \gamma} \frac{\mu}{z} \right) \left( \Pi_+ G_{\mu-\frac{1}{2}}(x,x')
+ \Pi_- G_{\mu+ \frac{1}{2}}(x,x')  \right)\ ,
\ee
which matches onto a similar result for fermions in AdS, after a Weyl rescaling \cite{Kawano:1999au}.
This Green's function imposes the boundary condition that $\Pi_+ G^F(x,x')|_{z \to 0} = 0$ along with the scaling constraint that $G^F \sim z^\mu$.  In the language of AdS/CFT, this scaling maps to a generalized free fermion propagating on the boundary with scaling dimension $\Delta = \mu + \frac{d-1}{2}$.  
We can also connect this result with the boundary OPE expansion in the context of boundary CFT.
Based on experience with the scalar field, we should be able to interpret $G^F(x,x')$ as the spinor boundary conformal block in a boundary OPE expansion.

We would like to compute the regulated value of $\langle \bar \psi e^{i \eta \gamma} \psi \rangle$ in the coincident limit. 
The point split expression is related to a trace of the Green's function, 
\be
\langle \bar \psi(x) e^{i \eta \gamma} \psi(x') \rangle &=& - \tr( e^{i \eta \gamma} G^F(x,x')) \\
&=& 
 \frac{{\mathfrak d}}{2} \left(\partial_z (G_{\mu+\frac{1}{2}}(x,x') - G_{\mu-\frac{1}{2}}(x,x')) + \frac{\mu}{z}(G_{\mu+\frac{1}{2}}(x,x') +G_{\mu-\frac{1}{2}}(x,x')) \right)\nonumber 
\ee
where ${\mathfrak d}$ is the rank of the Clifford algebra, nominally $\mathfrak{d} = 2^{\left\lfloor \frac{d}{2} \right\rfloor}$.

For the scalar, we had in the coincident limit that
$
\langle \phi^2 \rangle 
=  a_{\phi^2} z^{2-d}
$
where $a_{\phi^2}$ is given by (\ref{aphi2}).
Regulating the coincident limit by replacing $G_\mu(x,x')$ with $a_{\phi^2} (\mu) z^{2-d}$, we obtain
\be
\label{psibarpsitarget}
\langle \bar \psi e^{i \eta \gamma} \psi \rangle =  \frac{{\mathfrak d}}{4 z^{d-1}} \left( (2\mu-d+2) a_{\phi^2}(\mu+1/2) + (2\mu+d-2) a_{\phi^2}(\mu-1/2) \right) \ .
\ee

Assembling (\ref{dWdlambda}), (\ref{HSdimreg}), and (\ref{psibarpsitarget}), we can write an expression for the spinor hemisphere partition function:
\be
\label{Wspinor}
W_+(\mu) - W_+(0) = {\mathfrak d} \, \Gamma(1-d) \int_0^\mu \frac{\Gamma\left(\frac{d}{2} + \mu' \right)}{\Gamma\left(1 - \frac{d}{2} + \mu' \right) } \d \mu' \ .
\ee
The subscript $+$ indicates the $\Pi_+$ boundary condition.
As in the scalar case, we find a logarithmic divergence in every integer dimension, indicated by the $\Gamma(1-d)$ prefactor.
A similar expression for the partition function on AdS (which as we have argued is the same as the hemisphere partition function) was computed in the AdS/CFT literature in the context of double trace deformation in ref.\ \cite{Allais:2010qq}.
A more recent work \cite{Aros:2011iz} (see also \cite{Basar:2009rp,Aros:2010ng}) pointed out that this integral (\ref{Wspinor}) can be expressed as a Barnes' multiple 
gamma function:
\be
W_+(\mu) - W_+(0) \sim  {\mathfrak d} \log \Gamma_d \left( \frac{d}{2} + \mu \right) \ ,
\ee
``up to polynomial and logarithmic terms''.  We will come at this gamma function from a different angle, by explicitly diagonalizing the Dirac operator on the hemisphere in the presence of a conformal mass.

\subsection{The Hemisphere}
\label{sec:fermionHS}

Diagonalizing the Dirac operator on the hemisphere in the presence of a conformal mass is technically challenging.
We use a trick inspired by supersymmetry to relate the fermion eigenvalues and eigenfunctions to those of the scalar obtained in the previous section. With this method we are able to obtain the results for fermions in even dimensions 
with a slightly altered version of the mass term that involves the chirality matrix, $\bar \psi \gamma \psi$.

Our conventions for the fermion in curved space are that
$
\{ \gamma_\nu, \gamma_\rho \} = 2 g_{\nu\rho}
$
where $g_{\nu\rho}$ is the metric, $\gamma_\nu = e_\nu^a \gamma_a$, and $e_\nu^a$ are vielbein such that 
$\delta_{ab} e_\nu^a e_\rho^b = g_{\nu\rho}$.  The Lorentz generators are $\gamma^{ab} = \frac{1}{2}[\gamma^a, \gamma^b]$
leading to the Dirac operator
\be
\slashed{D} \psi 
= \gamma^\nu (\partial_\nu + \frac{1}{4} \omega_{ab\nu}\gamma^{ab} ) \psi \ ,
\ee
where $\omega_{ij\nu}$ is the spin connection. Additionally, the commutator of the covariant derivatives when acting upon a spinor is given by 
\begin{align} \label{}
[D_\mu, D_\nu] \psi = \frac{1}{4} R^{\rho\sigma}{}_{\mu\nu}  \gamma_{\rho\sigma} \psi = \frac{1}{2} \gamma_{\mu\nu} \psi \ ,
\end{align}
where in the last step we used the curvature of the sphere $R^{\rho\sigma}{}_{\mu\nu} = 2 \delta^{\rho}{}_{[\mu} \delta^{\sigma}{}_{\nu]}$. One finds the relation 
\begin{align} \label{}
\slashed{D}^2 = D^2 + \frac{1}{2} \gamma^{\mu\nu} [D_\mu, D_\nu] = D^2 - \frac{1}{4} d(d-1) \ .
\end{align}

As a warm up, let us first consider a fermion on a sphere (in any dimension) with no conformal mass.
We want to use Killing spinors $\epsilon$ to relate the eigenspinors $\psi$ of the Dirac operator and the eigenfunctions
$\phi$ of the scalar Laplacian.
The Killing spinor equation is \cite{Fujii:1985bg} 
\begin{align} \label{sphere_kse}
D_\mu \epsilon =  \frac{i}{2} \gamma_\mu \epsilon \ ,
\end{align}
which has $\mathfrak{d} = 2^{\left\lfloor \frac{d}{2} \right\rfloor}$ solutions.
Claim one is that if $\psi$ is an eigenspinor of the Dirac operator, $-i \slashed{D} \psi = \Lambda \psi$, then
$\bar \epsilon \psi$ is an eigenfunction of the scalar Laplacian.  A straightforward computation gives
\begin{align} \label{}
\left( D^2 - \frac{1}{4}d(d-2) \right) \bar \epsilon \psi 
&= - \Lambda ( \Lambda - 1) \bar\epsilon \psi \ . 
\end{align}
Comparing this result with the known eigenvalues of $D^2 - \frac{1}{4}d(d-2)$ which can be written in the form $- \lambda(\lambda - 1)$ with $\lambda = \ell + \frac{d}{2}$ and $\ell \in \mathbb{N}$ leads to $\Lambda = \lambda, -\lambda+1$. The possible eigenvalues are therefore $\pm ( \ell + \frac{d}{2})$. 

Claim two is that if $\phi$ is an eigenfunction of the scalar Laplacian, we can reconstruct the eigenspinors via
\begin{align} \label{sphere_egsp}
\psi_\pm
&=
\left(\left\{\!\begin{aligned}
&\lambda-1 \\
& -\lambda 
\end{aligned}\right\} - i \slashed{D} \right) \epsilon \phi \ ,
\end{align}
where $-i \slashed{D} \psi_+ = \lambda \psi_+$ and $-i \slashed{D} \psi_- = -(\lambda-1) \psi_-$.\footnote{%
We could just as well use the Killing spinor $\epsilon_-$ solving $D_\mu \epsilon_- = - \frac{i}{2} \gamma_\mu \epsilon_-$ with an extra sign relative to \eqref{sphere_kse} to write things more symmetrically as $\psi_\pm = ( \pm(\lambda-1) - i \slashed{D}) \epsilon_\pm \phi$ and $-i\slashed{D} \psi_\pm = \pm \lambda \psi_\pm$, but it will be more convenient for us below to work with a single type of Killing spinor.}
Let $\dim_S(\ell)$ be the degeneracy of the eigenvalue $\ell+ \frac{d}{2}$.\footnote{The degeneracy of the negative eigenvalues is the same as can be seen by changing $\gamma^\mu \to - \gamma^\mu$.}
Naively, this procedure gives an eigenspinor for each Killing spinor $\epsilon$ and each scalar eigenfunction $\phi$ of fixed $\ell$, which would amount to $\mathfrak{d}\times \dim(\ell)$ eigenspinors (see \eqref{scalar_deg}-\eqref{scalar_N_D} for the scalar degeneracy). However, not all the eigenspinors thus constructed are unique. Indeed, the operator $\lambda- 1 - i \slashed{D}$ in \eqref{sphere_egsp} has a non-trivial kernel; it vanishes on eigenspinors with eigenvalue $-(\lambda-1)$,  corresponding to $\ell-1$. Hence the kernel has dimension $\dim_S(\ell-1)$. This gives a recursion relation for the number of eigenspinors which leads to $\dim_S(\ell) = \mathfrak{d} \times \dim_N(\ell)$. (See \cite{Camporesi:1995fb} for a more standard derivation of these results.)

Now consider what happens when we add the conformal mass. The method suggested here relies on the relation between the scalar Laplacian and the square of the Dirac operator. The fermion operator we wish to diagonalize takes the form $-i \slashed{D} - i \frac{\mu}{\cos \theta}$. In Euclidean space the mass terms comes with an $i$ such that the full operator is not self-adjoint. Worse than that, due to the position dependence of the mass this fermion operator does not commute with its adjoint, and therefore they cannot be mutually diagonalized. What this means is that there is no simple relation between the fermion operator and the corresponding scalar Laplacian with a conformal mass. 

Nevertheless, in even dimensions the situation can be improved. Under a chiral transformation $\psi \to e^{i \eta \gamma/2} \psi$ the fermion operator transforms to
\begin{align} \label{}
e^{\frac{i\eta\gamma}{2}}  \left(-i \slashed{D} - i \frac{\mu}{\cos \theta} \right) e^{\frac{i \eta \gamma}{2}} 
=
\left(-i \slashed{D} - i e^{i \eta \gamma} \frac{\mu}{\cos \theta} \right) \ .
\end{align}
Let us first notice that the determinant of the fermion operator and therefore also the partition function does not change under this transformation. Moreover, for the choice $\eta = \pi/2$ the fermion operator becomes self-adjoint and we can use our method to find its eigenvalues and eigenfunctions.

We are thus led to consider the following eigenvalue problem
\begin{align} \label{}
\left(- i \slashed{D} + \gamma \frac{\mu}{\cos \theta} \right) \psi = \Lambda \psi \ .
\end{align}
To define the problem, we need to establish boundary conditions at the equator.  We will use analogs of the $\Pi_\pm$ that we introduced in flat space, and as a first step, it is useful to re-express these projection operators in terms of Killing spinors.

To write the Killing spinors in a useful way, it is useful to choose a basis.  The coordinates on the sphere are as above $(\theta_{1},\ldots, \theta_{d-1},\theta)$ with the usual metric (\ref{spheremetric}).
We take a basis for the gamma matrices given by
\begin{align} \label{}
\gamma^d = \begin{pmatrix} 0 & \mathds{1} \\ \mathds{1} & 0 \end{pmatrix} \ ,
\qquad
\gamma^a = \begin{pmatrix} 0 & -i \Gamma^a \\ i \Gamma^a & 0 \end{pmatrix} \ ,
\end{align}
where $\Gamma^a$ are the gamma matrices in one dimension less, and the chirality matrix is 
\begin{align} \label{}
\gamma = \begin{pmatrix} \mathds{1} & 0 \\ 0 & -\mathds{1} \end{pmatrix} \ .
\end{align}
In this basis we have
\begin{align} \label{}
\gamma^{ad} = -i \sigma^3 \otimes \Gamma^a \ , 
\qquad
\gamma^{ab} = \mathds{1} \otimes \Gamma^{ab} \ ,
\end{align}
where $a,b \neq d$.

The solution of the Killing spinor equation \eqref{sphere_kse} takes the following form \cite{Lu:1998nu}
\begin{align} \label{}
\epsilon = e^{\frac{i}{2} \theta \gamma^d} e^{- \frac{1}{2} \theta_{d-1} \gamma^{d-1,d}} \cdots e^{- \frac{1}{2} \theta_1 \gamma^{12}} \epsilon_0 \ ,
\end{align}
where $\epsilon_0$ is a constant spinor. In even dimension there are $2^{d/2}$ distinct choices of $\epsilon_0$ which correspond to different Killing spinors. We choose half of them in such a way that 
\begin{align} \label{sum_kills}
\sum \epsilon_0 \epsilon_0^\dagger = \begin{pmatrix} \mathds{1} &  \\  & 0 \end{pmatrix} \ .
\end{align}
We label this set of Killing spinors as $\epsilon_1$. 
Because all $\gamma^{ab}$ and $\gamma^{ad}$ are all block diagonal they commute through this matrix and we get
\begin{align} \label{}
\sum_{\epsilon_1} \epsilon \epsilon^\dagger 
= 
e^{\frac{i}{2} \theta \gamma^d} \begin{pmatrix} \mathds{1} &  \\  & 0 \end{pmatrix} e^{-\frac{i}{2} \theta \gamma^d}
&=
\frac{1}{2} \left( 1 + \cos \theta \sigma^3 + \sin \theta \sigma^2 \right) \otimes \mathds{1}
\nonumber \\
&=
\frac{1}{2} \left( 1 + \cos \theta \gamma + i \sin \theta \gamma^d \gamma \right) \ .
\end{align}
Choosing instead the other Killing spinors leads to
\begin{align} \label{}
\sum_{\epsilon_2} \epsilon \epsilon^\dagger 
= 
e^{\frac{i}{2} \theta \gamma^d} \begin{pmatrix} 0 &  \\  & \mathds{1} \end{pmatrix} e^{-\frac{i}{2} \theta \gamma^d}
=
\frac{1}{2} \left( 1 - \cos \theta \gamma - i \sin \theta \gamma^d \gamma \right) \ .
\end{align}
Overall, we get the relations
\begin{align} \label{}
\sum_{\epsilon_1} \epsilon \epsilon^\dagger |_{\theta =\pi/2} = \Pi_{-} \ ,
\qquad
\sum_{\epsilon_2} \epsilon \epsilon^\dagger |_{\theta =\pi/2} = \Pi_{+} \ .
\end{align}
It is important for verifying the formulas below that contrary to $\Pi_{\mp}$, the Killing spinor sum $\sum_{\epsilon_{1}} \epsilon \epsilon^\dagger$ annihilates $\epsilon_2$ at each point on the hemisphere, rather than just at the boundary (similarly with $\epsilon_1$ and $\epsilon_2$ exchanged).  
 
We are now finally in a position to go back to the fermion operator. Recall that the eigenvalues of the scalar with conformal mass on the sphere can be written as
\begin{align} \label{}
\left(D^2 - \frac{1}{4} d(d-2) - \frac{\mu(\mu-1)}{\cos^2 \theta} \right)\phi_{\mu-1/2} = - \lambda(\lambda-1) \phi_{\mu-1/2} \ ,
\end{align}
where $\lambda = \ell+\mu+\frac{d}{2}$ and similarly for $\phi_{\mu+1/2}$ with $\mu \to \mu+1$. Now suppose we have a eigenspinor $\psi$ with eigenvalue $\Lambda$. We can find the possible values of $\Lambda$ by constructing the scalar solutions 
\begin{align} \label{}
\phi_{\mu-1/2} = \bar\epsilon_1 \psi + c.c. \ , 
\qquad
\phi_{\mu+1/2} = \bar\epsilon_2 \psi + c.c. 
\end{align}
which satisfy the eigenvalue equation with masses $\mu(\mu-1)$ and $\mu(\mu+1)$ respectively with the same eigenvalue $- \Lambda(\Lambda-1)$. The possible eigenvalues are therefore $\pm( \ell + \mu + \frac{d}{2})$. Finally we can reconstruct the eigenspinors using the scalar solutions in the following way
\begin{align} \label{}
\psi_\pm 
&=
\left(\left\{\!\begin{aligned}
&\lambda-1\\
&-\lambda 
\end{aligned}\right\} + \frac{\mu \gamma}{\cos\theta} - i \slashed{D}\right) \epsilon_1 \phi_{\mu-1/2} \ . 
\end{align}
These spinors are annihilated by the projector $\Pi_+$ at the equator.  There are similarly also spinors constructed from $\epsilon_2 \phi_{-\mu-1/2}$ which are annihilated by $\Pi_-$ at the equator.  We can choose either set, depending on which boundary condition we prefer.  (On the sphere, we would require both sets.)

If we pick the boundary condition $\Pi_+ \psi = 0$, we find eigenvalues
$\lambda = \pm (\ell + \frac{d}{2} + \mu)$, each with degeneracy equal to that of the corresponding $\phi_{ \mu - 1/2}$ scalar field times the number $\mathfrak{d}/2$ of spinors $\epsilon_1$:
\be
\dim_{S}(\ell) \equiv {\ell +d - 1 \choose d-1 } \cdot \frac{\mathfrak{d}}{2} \ .
\ee
(Notice the factor of 1/2 compared to the result on the sphere.)
With the choice $\Pi_- \psi = 0$, the story is similar but with eigenvalues $\lambda = \pm (\ell + \frac{d}{2} - \mu)$.  
  Ignoring the sign of $\lambda$, we can immediately write the partition function up to a constant shift $c$ as
\be
W_\pm (\mu) = -{\mathfrak d} \sum_{\ell = 0}^\infty { \ell+d-1  \choose d-1} \log \left(\ell + \frac{d}{2} \pm \mu \right) + c \ ,
\ee
where the $\pm$ subscript indicates the choice of boundary conditions $\Pi_\pm$.  
Based on our earlier discussion of the Barnes' multiple gamma function, we identify 
\be
W_\pm (\mu) = {\mathfrak d} \, \log \Gamma_d \left( \frac{d}{2} \pm \mu \right)  + c \ .
\ee
The scalar partition function $W(\mu_s)$, subscript $s$ for scalar, 
thus becomes a sum of spinor partition functions with $\mu_s = \mu \pm \frac{1}{2}$.  

The $\Pi_+$ boundary condition is associated with a fermion of dimension $\Delta = \frac{d-1}{2} +\mu$ on the boundary, while the $\Pi_-$ condition links to $\Delta = \frac{d-1}{2} - \mu$.  Fixing the $\Pi_+$ boundary condition, the unitarity bound is reached at the boundary when $\mu=-\frac{1}{2}$.

\subsection{Computing the Partition Function}

The spinor partition function has a number of very similar features to the scalar one we considered earlier.  The integrand (\ref{Wspinor}) is an even/odd function of $\mu$ in odd/even $d$.  Thus the logarithmic contribution to $W_+(\mu) - W_+(-\mu)$ will vanish in even $d$.  
We can say more.  In odd $d$, the integrand is a polynomial in $\mu$ with no zeroes in the region $0<\mu<\frac{1}{2}$.  Hence, $W_+(\mu) - W_+(-\mu)$ is guaranteed to have a definite sign in this region and reach an extremum at the unitarity bound $\mu=\frac{1}{2}$. 
For example in $d=3$, we find 
\be
W_+(\mu) -W_+(-\mu) = \left( \frac{\mu}{4} - \frac{\mu^3}{3} \right) {\mathfrak d}  \log \frac{L}{\epsilon} \ .
\ee
which is positive in the range $0 < \mu < \frac{\sqrt{3}}{2}$ and reaches a maximum $\frac{{\mathfrak d}}{12} \log \frac{L}{\epsilon}$ at the unitarity bound $\mu=\frac{1}{2}$.  Setting ${\mathfrak d} = 2$, we obtain the usual result for the free fermion living on a 2d sphere.

In even $d$, we can try to isolate the constant term in $W_+(\mu) - W_+(-\mu)$ through zeta function regularization, as we did in the scalar case.  For example in 4d, we find that $W_+(\mu) - W_+(-\mu)$ is positive in the range $0<\mu<0.9004$ and reaches a maximum at the unitarity bound $\mu=\frac{1}{2}$ where
\be
W_+\left(\frac{1}{2} \right) - W_+\left(-\frac{1}{2} \right) = \frac{1}{8} \left( \log(2) + \frac{3 \zeta(3)}{2 \pi^2} \right) \ ,
\ee
using ${\mathfrak d} = 4$, corresponding to the free energy of a fermion on an $S^3$ \cite{Klebanov:2011gs}.  
We could pursue these calculations in other $d$, but they are analogous to calculations in for example \cite{Allais:2010qq,Aros:2011iz} in the context of AdS/CFT and we shall again spare the reader.

\section{Discussion}

We started with the observation that there is potentially a large class of unexplored marginal deformations in boundary CFT of the form (\ref{non_marginal_def}).  Given an operator ${\mathcal O}$ with a protected scaling dimension, we can deform the theory with a position dependent coupling that preserves the underlying $SO(d,1)$ conformal symmetry group.  We then asked how the boundary CFT behaves with respect to such deformations.  While the answer to the general question is difficult, we were able to make some progress.  Using conformal perturbation theory, we can express the derivative of the partition function with respect to the deformation parameter (\ref{HSdimreg}) in terms of the one-point function coefficient $a_{\mathcal O}$ where $\langle {\mathcal O} \rangle = a_{\mathcal O} z^{-\Delta}$.  Furthermore, in the case of free scalars and fermions, we were able to compute $a_{\phi^2}$ and $a_{\bar \psi \psi}$ exactly, integrate the derivative, and express the partition function globally as a function of the deformation parameter.

In fact, due to a Weyl scaling argument, we saw that this partition function 
computation maps onto analogous calculations of the partition function of massive fermions and scalars in anti-de Sitter space.  These partition functions have a long history, a recent incarnation of which is in the discussion of double trace deformations \cite{Gubser:2002vv,Allais:2010qq} in AdS/CFT correspondence. 
While the determination of the partition function by integrating $a_{\mathcal O}$ has essentially been done before in the AdS/CFT context, we believe calculations 
in sections \ref{sec:scalarHS} and \ref{sec:fermionHS}
of the partition function by diagonalizing the equations of motion on the hemisphere in the presence of a conformal mass are new.
This alternate pathway has the benefit of quickly showing the relation between the effective action and the Barnes' multiple gamma function.

Despite the absence of interactions, the free fermion and scalar have much to offer.  Alternately through the boundary OPE or through the AdS/CFT dictionary, we can interpret the physics living on the boundary of these theories to be that of generalized free scalar or fermionic fields.  The partition function in the bulk thus affords a way of computing the determinant of the fractional Laplacians and Dirac operators that govern the behavior of the generalized free fields on the boundary \cite{Kwasnicki}.  Moreover, in the unitarity bound limit $\mu = -1$ for the scalar and $\mu= -\frac{1}{2}$ for the fermion,
we recover exactly free fields on the boundary.

The link to AdS/CFT suggests a way of possibly realizing AdS/CFT in the laboratory.  While constructing anti-de Sitter space experimentally seems prohibitively difficult, realizing a set-up in flat space with position dependent couplings that fall off as we move away from a boundary seems more feasible.  In the case of marginal deformations, for example the $F_{\mu\nu} F^{\mu\nu}$ example discussed in the introduction in connection with a mixed dimensional QED, 
the couplings can even be position independent.  Indeed, 
the mixed dimensional QED theory is similar in some respects to graphene and has been proposed as the ultimate IR fixed point as the speed of the electron grows at longer distances \cite{Teber:2018goo}.

An issue which deserves further clarification is the precise relation between BCFT with (\ref{non_marginal_def}) 
and field theory in AdS.  We argued 
at the classical level that these two field theories are equivalent by Weyl rescaling.  
Generically 
renormalization group effects should 
 spoil the marginality of the deformation (\ref{non_marginal_def}) and hence break conformal symmetry in flat space.  
 In AdS, on the other hand, the symmetry group is an isometry of the space-time and likely to be more robust.  
 At the quantum level, the difference
between flat space BCFT and field theory in AdS should come from how the theory is regulated.

One intriguing direction for future research is the possibility of having time dependent couplings.  While we focused on the Euclidean setting, in the Lorentzian setting it is well known that the $Y^4$ theory we studied in section \ref{sec:emergence} can support a profile 
$Y \sim 1/ t$ which preserves the same symmetries as de Sitter space.  Some researchers have already investigated implications for cosmology (see e.g.\ \cite{Coradeschi:2013gda}).

\section*{Acknowledgments}
We would like to thank O.~Aharony, D.~Anninos, F.~Benini, N.~Doroud,  Z.~Komargodski, E.~Lauria, and M.~Serone for discussion.
This work was supported in part by the U.S.\ National Science Foundation Grant PHY-1620628, 
by the U.K.\ Science \& Technology Facilities Council Grant ST/P000258/1, by the
ERC Starting Grant N.~304806.  C.~H. was supported in part by a Wolfson Fellowship from the Royal Society.
I.S. is supported in part by the MIUR-SIR grant RBSI1471GJ ``Quantum
Field Theories at Strong Coupling: Exact Computations and
Applications'' and by INFN Iniziativa Specifica ST\&FI.

\appendix

\section{Maps to Other Examples} 
\label{app:maps}

For the spherical boundary in ${\mathbb R}^d$, 
consider the sphere $(z-L)^2 + \delta_{ij} x^i x^j = L^2$ centered at $z=L$ and $x^i = 0$. We use an inversion that does not preserve the boundary, namely one about the origin of the coordinate system, $z \to \frac{z}{z^2 + \delta_{ij} x^i x^j}$ and $x^i \to \frac{x^i}{z^2 + \delta_{ij} x^i x^j}$, mapping the sphere to a plane at $z = \frac{1}{2L}$.  The Weyl factor is $\Omega(x) = \frac{1}{z^2 + x^i x^j \delta_{ij}}$.  
The distance $z-\frac{1}{2L}$ to the planar boundary becomes the function $\frac{1}{2L} ( \delta_{ij} x^i x^j + (z-L)^2  - L^2)$ in the frame with a spherical boundary.  For simplicity, we can then recenter the plane and the sphere at $z=0$, using translations.  The distance $z$ from the plane becomes the factor $\frac{1}{2L} (r^2 - L^2)$ (using $r$ for the distance from the origin) in the system with a spherical boundary, allowing us to write a perturbation of the form (\ref{balldef}).

For the hemisphere, it is convenient to start from the case of a spherical rather than a planar boundary in ${\mathbb R}^d$.  We transform an $S^d$ of radius $L$ to flat space, but where the equator $\theta = \frac{\pi}{2}$ of the $S^d$ is mapped to a $S^{d-1}$ of radius $r=L$ in ${\mathbb R}^d$.  
We have
\be
\d s^2 &=&  L^2 ( \d \theta^2 + \sin^2 \theta \, \d \Omega_{d-1}^2 )  \nonumber \\
&=& \Omega^2(x) \left( \d r^2 + r^2 \, \d \Omega_{d-1}^2 \right) \ ,
\ee
with $L \tan \frac{\theta}{2} = r$ and $r \Omega  = L \sin \theta$.  The distance to the plane $z$ becomes the factor
 $L \cos \theta$ on the hemisphere, leading to a perturbation of the form (\ref{HSdef}).  

\section{A Particular Integral}
\label{app:tricks}

To extract the log divergence from the quantity $W(\mu) - W(0)$ in the text, we need to be able to evaluate the following integral:
\be
I(a,b,\mu) &=& \frac{1}{2} \int_0^{\ell_{\rm max}} {a + \ell \choose a} \log \left( 1 + \frac{\mu}{\ell + b} \right) \d \ell \\
&=&  \frac{1}{2 \Gamma(a+1)}  \int_0^{\ell_{\rm max}} (\ell+a) (\ell+a-1) \cdots (\ell+1) \log \left( 1 + \frac{\mu}{\ell+b} \right) \d \ell \ .
\nonumber 
\ee
We further introduce
\be
I(a,b,\mu,y) &=& \frac{1}{2 \Gamma(a+1)} \int_0^{\ell_{\rm max}} (\ell+a) (\ell+a-1) \cdots (\ell+1) \log \left(1 + \frac{\mu}{\ell+b} \right) y^\ell \d \ell \nonumber \\
&=& \frac{1}{2 \Gamma(a+1)} \frac{\partial^a}{\partial y^a} \int_0^{\ell_{\rm max}} \log \left(1 + \frac{\mu}{\ell+b} \right) y^{\ell + a} \d \ell \ , 
\ee
and study instead the $\mu$ derivative 
\be
\partial_\mu I(a,b,\mu,y) &=&  \frac{1}{2 \Gamma(a+1)} \frac{\partial^a}{\partial y^a} \int_0^{\ell_{\rm max}} \frac{y^{\ell+a}}{\ell + b + \mu} \d \ell
\ee
which integrates to an exponential function.  
The $\log \ell_{\rm max}$ contribution is then straightforward to isolate:
\be
\partial_\mu I(a,b,\mu,1) &\sim& \frac{1}{2 \Gamma(a+1)} \lim_{y\to 1} \frac{\partial^a}{\partial y^a} y^{a - b - \mu} \log(\ell_{\rm max}) \nonumber \\
&=& \frac{1}{2} {a-b-\mu \choose a} \log(\ell_{\rm max}) \ .
\ee

\bibliographystyle{JHEP}
\bibliography{conformalmass_bib}

\end{document}